 \documentclass[twocolumn,aps,prb,superscriptaddress]{revtex4-2}
\usepackage{amsmath,empheq}
\usepackage{amssymb}
\usepackage{mathrsfs}  
\usepackage{amsfonts}
\usepackage{booktabs}
\usepackage{physics}
\usepackage{graphicx} 
\usepackage{subfigure} 
\usepackage{epsfig}
\usepackage{color}
\usepackage{epstopdf}
\usepackage{rotating}
\usepackage{appendix}
\usepackage{chemformula}
\usepackage{bm}
\usepackage{makecell}
\usepackage[utf8]{inputenc}
\usepackage[colorlinks=true,linkcolor=blue,citecolor=blue,filecolor=blue,urlcolor=blue]{hyperref}
\usepackage[export]{adjustbox}

\begin{document}

 \title{Topological superconductivity from doping a triplet quantum spin liquid in a flat band system}
 \author{Manuel Fern\'andez L\'opez}
\affiliation{Departamento de F\'isica Te\'orica de la Materia Condensada, Condensed Matter Physics Center (IFIMAC) and
Instituto Nicol\'as Cabrera, Universidad Aut\'onoma de Madrid, Madrid 28049, Spain}
\author{Ben J. Powell}
\affiliation{School of Mathematics and Physics, The University of Queensland, QLD 4072, Australia}
\author{Jaime Merino}
\affiliation{Departamento de F\'isica Te\'orica de la Materia Condensada, Condensed Matter Physics Center (IFIMAC) and
Instituto Nicol\'as Cabrera, Universidad Aut\'onoma de Madrid, Madrid 28049, Spain}

\begin{abstract}
We explore superconductivity in   
strongly interacting electrons on a decorated honeycomb lattice (DHL). 
An easy-plane ferromagnetic  interaction arises from spin-orbit coupling in the Mott insulating phase, which favors a triplet resonance valence bond  spin liquid state.
Hole doping leads to partial occupation of a flat band and to triplet superconductivity.
The order parameter is highly sensitive to the doping level and the interaction parameters, with $p+ip$, $f$ and $p+f$ superconductivity found, as the flat band leads to instabilities in multiple channels.
Typically, first order transitions  separate  different superconducting phases, but a second order transition separates two time reversal symmetry breaking $p+ip$ phases with different Chern numbers ($\nu=0$ and 1). 
The  Majorana edge modes in the topological  ($\nu=1$) superconductor are almost localized due to the strong electronic correlations  in a system with a
flat band at the Fermi level. This suggests that these modes could be useful for topological quantum computing. 
The `hybrid'  $p+f$ state does not require two phase transitions as temperature is lowered. This is because the symmetry of the model is lowered in the $p$-wave phase, allowing arbitrary admixtures of $f$-wave basis functions as overtones.
We show that the multiple sites per unit cell of the DHL, and hence multiple bands near the Fermi energy, lead to very different nodal structures in real and reciprocal space. We emphasize that this should be a generic feature of multi-site/multi-band superconductors.
%Hybrid superconductivity involving combined $p$ and $f$ superconducting orders
%is favored in the isotropic limit of the DHL. In contrast, at sufficiently large geometrical anisotropy, we find a region of topological superconductivity that breaks time reversal symmetry. Our results are relevant to ongoing effects to realise superconductivity in organic and organometallic compounds, such as Rb$_3$TT$\cdot$2H$_2$O and Mo$_3$S$_7$(dmit)$_3$, coordination polymers, metal organic frameworks, and covalent organic frameworks.
\end{abstract}
 \date{\today}

 \maketitle
 
\section{Introduction}

Understanding the mechanism of superconductivity in strongly correlated 
materials remains a formidable challenge. For instance, high-$T_c$ cuprates \cite{Norman2005,Lee2006,Ogata2008}, organics\cite{Powell2011} and twisted bilayer graphene \cite{Cao2018,Jarillo2018} (TBG), display similar phase diagrams\cite{Mckenzie1997}. In these materials superconductivity emerges near to a Mott insulating phase, 
indicating the important role played by strong correlations on Cooper pairing.
A common ingredient in these systems is the presence of 
antiferromagnetic interactions, which  lead to  Mott insulators with antiferromagnetic order, as in the cuprates, or quantum spin liquids, as in the organic materials 
$\kappa$-(BEDT-TTF)$_2$Cu(CN)$_3$ or $\kappa$-(BEDT-TTF)$_2$Ag(CN)$_3$
\cite{Shimizu2005,Shimizu2016,Powell2011}.  Such AF interaction
is crucial to singlet $d$-wave (or $d+id$) superconductivity which, according to Andersons
theory of high-$T_c$ superconductivity \cite{Anderson1987}, can emerge 
when hole doping the resonance valence bond (RVB) Mott insulator on the square (triangular \cite{Powell2007}) lattice \cite{Anderson1973}.  
The discovery of superconductivity 
in magic angle TBG \cite{Jarillo2018} poses new questions about strongly correlated superconductivity in flat bands   \cite{Balents}. 

Definitive signatures of triplet pairing are rare outside the well understood triplet superfluidity \cite{Leggett2004,Leggett1975} in $^3$He. At present, there is no unambiguous 
evidence for, and increasing evidence against, triplet pairing in (former) candidate materials 
such as Sr$_2$RuO$_4$ \cite{MacKenzie2017,MacKenzie2003,Pustogow}, the quasi-one-dimensional (TMTSF)$_2$X Bechgaard salts\cite{Wosnitza2019}
or A$_2$Cr$_3$As$_3$ \cite{Bao2015,Yang2021}, (A=K, Rb, Cs).   A reason for the scarcity of triplet superconductors is 
the  antiferromagnetic (AFM) superexchange between spins
which often dominates over   ferromagnetic exchange processes, likely to stabilize triplet superconductivity. However, the superconductivity observed \cite{Cai2020} in the ferromagnetic Mott insulators, CrXTe$_3$ with $X=$Si, Ge has been predicted to be of the triplet type \cite{Konig2022}. Perhaps the most promising class of materials for observing triplet superconductors are the uranium based heavy fermion materials \cite{Jiao,Aoki}, where superconductivity is often found near ferromagnetism.  

Quantum spin liquids with ferromagnetic interactions cannot be described through standard singlet RVB theory. 
However, recent extensions to easy-plane
ferromagnetic triangular lattices predict the existence of triplet resonance valence bond (tRVB)  
Mott insulators which can become unconventional $p+ip$-wave superconductors under hole doping\cite{Konig2022}. On the other hand, in multiorbital systems such as the iron pnictides, Hunds coupling can induce an intra-atomic triplet RVB state \cite{Coleman2020}. In these cases, triplet superconductivity may arise under hole doping. This is
 allowed by the presence of an even number of atoms per unit cell, leading to a spatially staggered gap pattern, as proposed by Anderson \cite{Anderson1985} in the context of heavy fermion superconductivity. 

Triplet pairing induced by ferromagnetic interactions may
arise in certain organic and organometallic materials with 
unit cells containing many atoms. (EDT-TTF-CONH$_2$)$_6$[Re$_6$Se$_8$(CN)$_6$] \cite{Baudron},
Mo$_3$S$_7$(dmit)$_3$ \cite{Llusar2004}, and Rb$_3$TT$\cdot$2H$_2$O \cite{Shuku2018} crystals  with
layers of decorated honeycomb lattices (DHLs) can potentially host
rich physics arising from the interplay of strong correlations, Dirac points, quadratic band touching points and flat bands  \cite{Manuel2020,Powell2021A,Powell2021B,Manuel2022,Nourse22}. This lattice also occurs in several metal organic frameworks and coordination polymers \cite{Nourse22}. Indeed, unconventional {\it singlet} $f$-wave pairing has been found in an AFM $t$-$J$ model on the 
DHL \cite{Merino2021}. However, the spin molecular-orbital coupling (SMOC) present in these systems,\cite{Khosla2017,Jacko2017,Merino2017,Powell2017} can lead to easy-plane ferromagnetic interactions favoring tRVB states.
Hence, it is interesting to address the question of whether triplet 
superconductivity emerges in a strongly correlated easy-plane ferromagnetic
model on a DHL. The possibility of finding non-trivial topological superconductivity as well as the role played by the flat bands 
deserve special attention. 

In this paper we study a   $t$-$J$ 
model on a DHL with XXZ interactions, which arise due to spin-orbit coupling. Exact diagonalization of small clusters shows that a  tRVB spin liquid,  is a competitive ground state of the model at half-filling.
Hole doping this spin liquid state leads to partial occupation of a flat band.
Based on this we apply the tRVB approach to search for superconductivity in the hole doped system. We find multiple  triplet superconducting phases, including include %$p$
$p+ip$, $p+f$, $f$. Presumably, the wide variety of superconducting phases is a consequence of the flat band leading to instabilities in multiple Cooper channels. 

Interestingly, we find both topologically trivial and topological $p+ip$ superconductivity with Chern numbers, $\nu=0$ and 1 respectively. These phases are separated by a continuous phase transition, where nodes appear in the otherwise fully gapped $p+ip$ order parameter. In the topological superconductor phase  the single Majorana edge mode expected for $\nu=1$ is almost localized since it traverses a tiny gap bounded by nearly  flat bands. 

We show that the $p+f$ state does not require a two superconducting phase transitions as the temperature is lowered. Rather the once the symmetry of the system is lowered by going into the $p$-wave superconducting phase, the $p$ and $f$ solutions belong to the same irreducible representation of the group describing the symmetry of the model. Therefore, $p$ and $f$ are overtones and the system can and does take advantage of this to lower its energy.

The DHL lattice has six sites per unit cell. Therefore, diagonalizing the Hamiltonian requires a Bogoliubov transformation,  a Fourier transformation and a transformation from the multi-site basis to a multiband basis. We show that the latter leads to remarkably different nodal structures in real and reciprocal space. We emphasize that this is a very general phenomenon in multi-band superconductors.

We show that, at strong coupling (low doping), where the magnetic exchange is dominant, real space pairing dominates, leading to  order parameters that are fully gapped or have only isolated nodal points in reciprocal space. In contrast,  at weak coupling (high doping), where the superconductivity is dominated by the kinetic energy, 
the superconducting gap displays 
nodal points in reciprocal space. 

The paper is organized as follows: in Sec \ref{sec::model} we introduce the anisotropic $t$-$J$ model on a DHL with XXZ magnetic exchange interactions. 
In Sec. \ref{sec::TRVBQSL} we show, using exact diagonalization of small clusters, that a tRVB spin liquid,  is a competitive ground state of the model at half-filling. 
In Sec. \ref{sec::TRVBSC} we show that a wide variety triplet superconducting states emerge on hole-doping the tRVB, including a topological superconductor (TSC), and give a detailed characterization of these states. 
In Sec. \ref{sec::TSC} we characterize the TSC by calculating the topological invariants and Majorana edge modes.  
Finally, in Sec. \ref{sec::conclusion} we conclude our work by summarizing our main results
and their relevance to actual materials realizing DHLs.

\section{Model}
\label{sec::model}

\begin{figure}
\centering
\includegraphics[width=8cm]{ 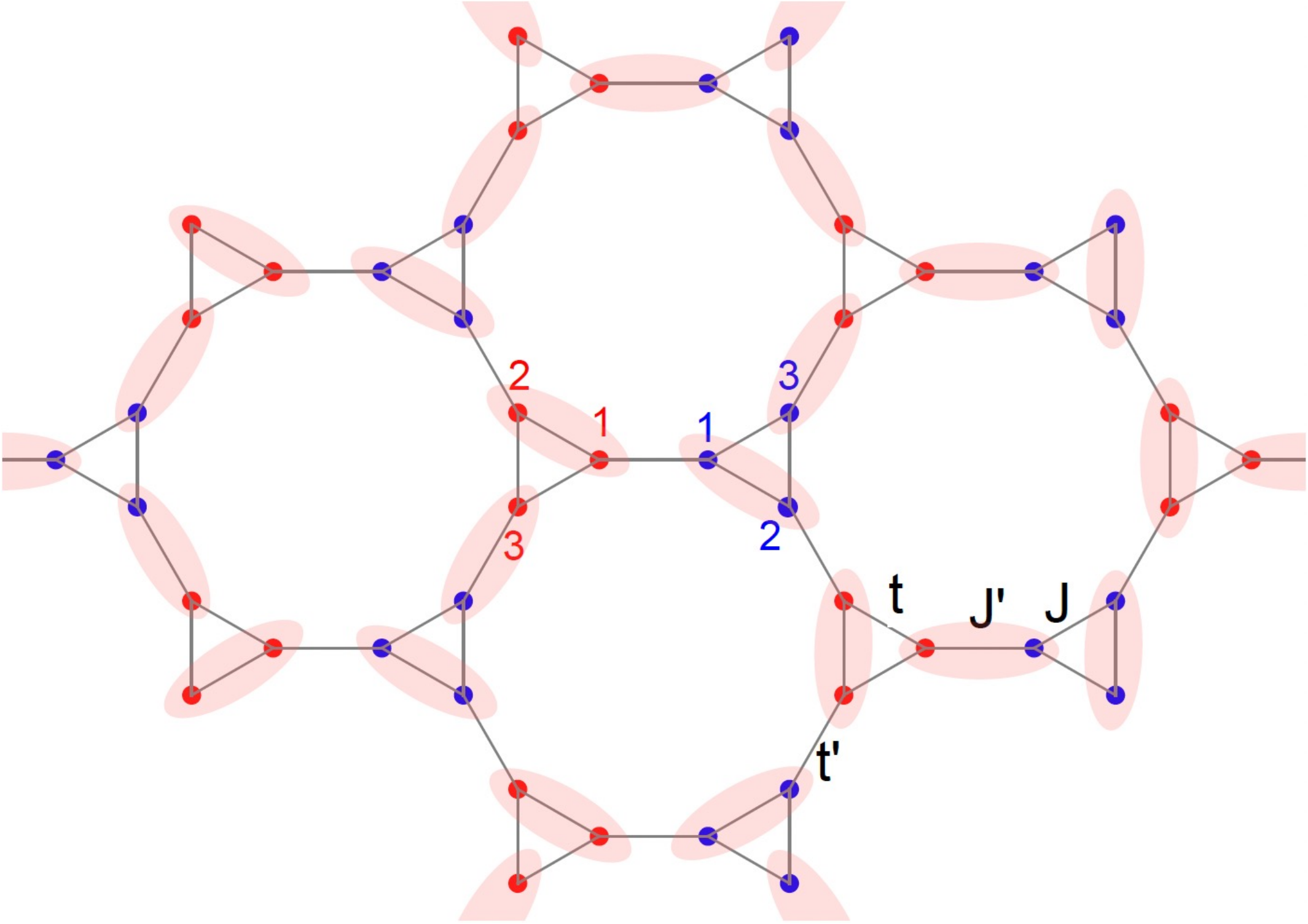}
\caption{The $t$-$J$ model on the decorated honeycomb lattice. The unit cell consists of a A (red) and a B (blue) triangles with the sites  numbered as shown. The parameters entering the model are illustrated. A particular triplet covering of the lattice entering the $| \text{tRVB} \rangle$ state of the model is shown.}
\label{fig:cluster}
\end{figure}

Our starting point is a $t$-$J$ model with easy-plane XXZ ferromagnetic exchange on the DHL:
\begin{eqnarray}
{\cal H} &=& -t \sum_{ \alpha i  j  \sigma} P_G \left( c^\dagger_{\alpha i \sigma} c_{\alpha j, \sigma}  + c^\dagger_{\alpha j \sigma} c_{\alpha i \sigma} \right) P_G
\nonumber \\
&-& t' \sum_{\substack{\langle A, B  \rangle i \sigma }} P_G \left( c^\dagger_{A i \sigma} c_{B i\sigma}  + c^\dagger_{B i \sigma} c_{A i \sigma} \right) P_G  \nonumber \\
&-& J \sum_{ \alpha i j  } \left( S^x_{\alpha i}  S^x_{\alpha j} +S^y_{\alpha i}  S^y_{\alpha j} - S^z_{\alpha i}  S^z_{\alpha j} + {\frac{1}{4}} n_{\alpha i} n_{\alpha j} \right) 
\nonumber \\
&-& J'\sum_{\substack{\langle A , B  \rangle i }} \left( S^x_{A i}  S^x_{B i} + S^y_{A i}  S^y_{B i} -S^z_{A i}  S^z_{B i} + {\frac{1}{4}} n_{A i} n_{B i} \right) \nonumber \\
&+& \mu \sum_{\alpha i \sigma} c^\dagger_{\alpha i \sigma} c_{\alpha i \sigma},
\label{eq:modeltJ}
\end{eqnarray}
where $P_G=\Pi_i (1- n_{i\uparrow} n_{i\downarrow} )$ is the Gutzwiller projection operator which excludes 
doubly occupied sites completely,  $c^{(\dagger)}_{\alpha i \sigma}$ is the usual annihilation (creation) operator, and $S^r_{\alpha i}$ is the  $r$th component of the spin operator. The $\alpha$-index runs over the two triangular clusters, while $i,j$ run 
over the three sites within each triangular clusters, with
site numbering as illustrated in Fig. \ref{fig:cluster}, and $\sigma$ denotes the spin of the electron.
Angled brackets indicate that the sums are restricted to (triangles that contain) nearest-neighbor  sites. 
We are interested in the properties of the model close to half-filling, so we write the electron density 
$n=1-\delta$, where $\delta$ is the density of holes doped into the half-filled DHL. We fix $t=1$ 
as the energy scale. The XXZ exchange couplings, $J,J'>0$, lead to easy-plane ferromagnetic (FM)  interactions
and AFM longitudinal interaction. 

A possible microscopic origin of the  XXZ exchange 
interactions of model (\ref{eq:modeltJ}) can be found by considering a single s-like orbital with on site  
(Hubbard) interactions and SMOC. On transforming to real space SMOC is equivalent to an anisotropic Kane-Mele  spin-orbit coupling. If we consider  nearest neighbor (nn) interactions with a $z$ component of the spin-orbit coupling only we have
\begin{equation}
    \mathcal{H}=i\lambda\sum_{\langle \alpha i\beta j\rangle}(c_{\alpha i\uparrow}^\dagger c_{\beta j\uparrow}-c_{\alpha i\downarrow}^\dagger c_{\beta j\downarrow})+U\sum_{i\alpha}n_{i\alpha\uparrow}n_{i\alpha\downarrow},
    \label{eq:Hubbard}
\end{equation}
where $n_{i\alpha\sigma}=c_{i\alpha\sigma}^\dagger c_{i\alpha\sigma}$. In the large-$U$ limit, the Hubbard model maps onto an effective  
XXZ spin hamiltonian \cite{LeHur2010}
\begin{equation}
   \mathcal{H}_\text{XXZ}= -{ \frac{4\lambda^2}{U}} \sum_{\langle \alpha i\beta j\rangle} \left( S^x_{\alpha i}  S^x_{\beta j} +S^y_{\alpha i}  S^y_{\beta j} - S^z_{\alpha i}  S^z_{\beta j} \right).
   \label{eq:xxz}
\end{equation}
At half-filling, the Hubbard model (\ref{eq:Hubbard}) with 
$U\gg |\lambda|$ becomes an XXZ 
model with $J=4 \lambda^2/U, J'= 4 \lambda'^2/U$, i.e. Eq. \eqref{eq:modeltJ}, as is corroborated by our exact denationalization (ED) calculations of Appendix \ref{app::pairingED}.
On including direct the hopping terms $t$ and $t'$ additional AFM exchange, Dzyaloshinskii-Moriya, and off diagonal exchange interactions will be introduced \cite{Khosla2017,Trebst,Winter}. 
We neglect these interactions for simplicity. Thus Eq. (\ref{eq:modeltJ}) is a toy model for understanding the impact of SMOC on superconductivity.  We further assume that $\lambda \propto t$ and $\lambda' \propto t'$ so that $J'/J = (t'/t)^2$ reducing the number of 
independent parameters in our Eq. \eqref{eq:modeltJ}.
Other mechanisms for easy-plane ferromagnetic interaction have also been put forward recently, including interactions arising from the interplay of Hund's and Kondo interactions in heavy fermions.\cite{Coleman2022} 

We emphasize that different materials are known to span a wide range of the parameter space for this model. For example, in Mo$_3$S$_7$(dmit)$_3$ is in the trimerized limit: $t'<t$ and $J'<J$ \cite{Jacko2015,Jacko2017}, whereas in (EDT-TTF-CONH$_2$)$_6$[Re$_6$Se$_8$(CN)$_6$] \cite{Baudron} and Rb$_3$TT$\cdot$2H$_2$O  \cite{Shuku2018} are in the dimerized limit: $t'>t$ and $J'>J$. The physics of these two limits is known to differ markedly \cite{Powell2021A,Nourse22}. Furthermore, at ambient pressure, (EDT-TTF-CONH$_2$)$_6$[Re$_6$Se$_8$(CN)$_6$] undergoes a metal-insulator transition as the temperature is lowered \cite{Baudron}, whereas Mo$_3$S$_7$(dmit)$_3$ \cite{Llusar2004} and Rb$_3$TT$\cdot$2H$_2$O  \cite{Shuku2018} are insulating at all temperatures at ambient pressure.

\section{Triplet RVB quantum spin liquid}
\label{sec::TRVBQSL}
We analyze the ground state of model (\ref{eq:modeltJ}) 
by using tRVB theory\cite{Momoi2009}, which has
recently applied to iron pnictides\cite{Coleman2020} and transition metal chalcogenides\cite{Konig2022}.  
It is  analogous to Andersons RVB theory of high-$T_c$ superconductivity to deal with strongly correlated models containing ferromagnetic exchange. Triplets rather than singlet bonds are the building blocks of the theory. The simplest form of the theory assumes that the ground state 
of the ferromagnetic undoped model is a quantum spin liquid in which spins 
are paired  into triplets. 

In general,
such triplet RVB state can be expressed as:
\begin{equation}
|\text{tRVB} \rangle = \sum_P A_P |P_t \rangle, 
\end{equation}
where $|P_t \rangle = \Pi_{{ij} \in P} |ij\rangle$ and
the  $S_z=0$   spin triplet between sites $i, j$ is
\begin{equation}
    |ij\rangle \equiv \frac{|i,\uparrow \rangle |j, \downarrow \rangle + |i,\downarrow \rangle |j, \uparrow \rangle} {\sqrt{2} },
\end{equation}
where $|ij\rangle=-|ji\rangle$ due to the symmetry of triplets under inversion. Hence, the $|t\text{RVB}\rangle$ state is a quantum superposition of all possible coverings of the lattice into triplet valence bonds. The simplest version of the $|t\text{RVB}\rangle$ state involves triplets between nn spins only. One possible snapshot triplet configuration $|P_t \rangle $ entering $|t\text{RVB}\rangle$ is shown in Fig. \ref{fig:cluster}. 
\begin{figure}
\centering
\includegraphics[width=8cm]{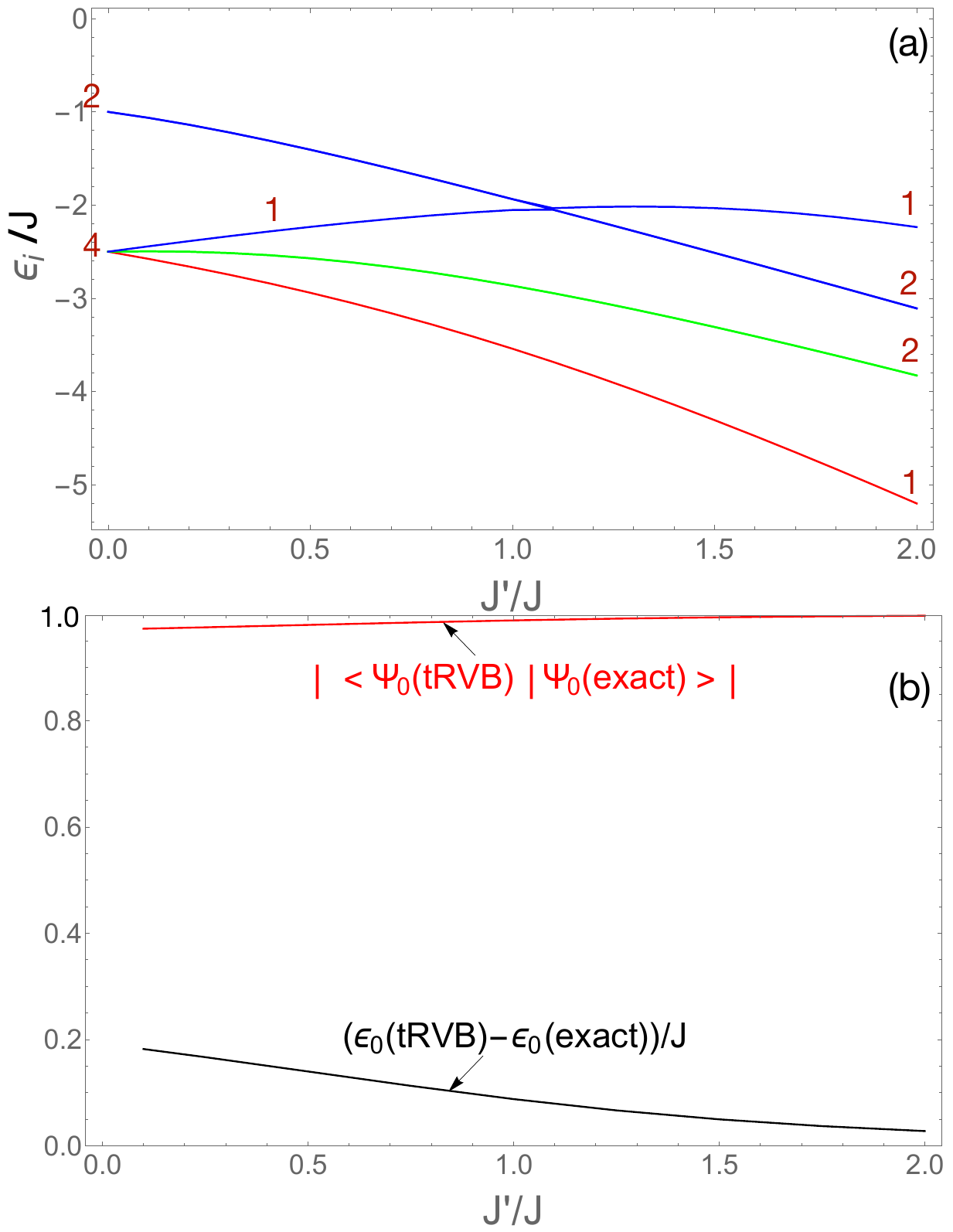}
% RVB_comparison.pdf}
\caption{Exact triplet RVB analysis of the half-filled easy-axis XXZ ferromagnetic model on a six site (two triangle) cluster.
 (a) The dependence of the low energy  spectrum of the model \eqref{eq:modeltJ} on $J'/J$. The numbers denote the exact energy level degeneracies.  (b) The overlap between the exact ground state and the tRVB wavefunction for the six site cluster (given in Eq. (\ref{eq:varRVB})), $|\langle \Psi_0(\text{tRVB}) | \Psi_0 (\text{exact})\rangle |$, and the difference between the exact ground state energy, $\epsilon_0(\text{exact})$,  and the triplet RVB energy, $\epsilon_0(\text{tRVB})$. 
%Energies are given in units of $J=1$. 
}
\label{fig:fig2}
\end{figure} 
%The exchange coupling contribution to the model can be decoupled exactly in terms of triplet 
%valence bonds:
%\begin{equation}
%H_J= -J \sum_{\langle ij \rangle} (S_i^x S_j ^x + S_i^y S_j^y - S_i^z S_j^z + {1 \over 4} n_i n_j)= -J \sum_{\langle ij \rangle} h^\dagger_{ij} h_{ij},
%\end{equation}
%where:
%\begin{equation}
%h^+_{ij}={1 \over \sqrt 2} (c^\dagger_{i \uparrow} c^\dagger_{j\downarrow} +c^\dagger_{i \downarrow} c^\dagger_{j\uparrow}  ),
%\end{equation}
%creates a triplet ($S=1$, $S_z=0$) valence bond between electrons located at $i$ and $j$. 

\subsection{Exact analysis of the triplet RVB in the DHL}
\label{sec::exacttriplet}
We first concentrate on model (\ref{eq:modeltJ}) at half-filling, $\delta=0$, which becomes a easy-plane XXZ ferromagnetic model.  
Triplet RVB states in such spin model
are explored by performing exact diagonalization on six-site (two-triangle) clusters. The dependence of the exact level spectra of the cluster with $J'/J$ 
is shown in Fig. \ref{fig:fig2}(a). For any $J'/J \neq 0$ the four-fold 
ground state degeneracy found at $J'=0$ is split so that the ground state  
becomes a non-degenerate $S=1, S_z=0$ triplet, as expected. The first excitation
corresponds to a $S=1, S_z=\pm 1$ doublet. The splitting between the 
$S=1, S_z=0$ and $S=1,S_z=\pm1$ states is increased by $J'/J$. 
The ground state energy at $J'=J$ is $\epsilon_0(\text {exact})=-3.541287J$, 
and the corresponding wavefunction is given in the Appendix. \ref{app::EDanalysis}.

We now consider a simple nn tRVB ansatz for the ground state wavefunction 
of the cluster, which consists on a linear combination of nn triplet valence bonds (VB) only:
\small
\begin{eqnarray}
%|\text{nn-tRVB} \rangle &=&|14\rangle |23\rangle |56\rangle +|13\rangle |25\rangle |46\rangle 
%\nonumber \\
%&&+ |12\rangle |36\rangle |45\rangle -|14\rangle |25\rangle |36\rangle .
|\text{tRVB} \rangle &=&  |{A1B1}\rangle |{A2A3}\rangle |{B2B3}\rangle  + |{A1A3}\rangle |{A2B2}\rangle |{B1B3}\rangle
\nonumber\\
&&+|{A1A2}\rangle |A3B3\rangle |B1B2\rangle 
-|A1B1\rangle |{A2B2}\rangle |{A3B3}\rangle ,
\nonumber\\\normalsize
\label{eq:tRVB}
\end{eqnarray}
with the corresponding energy
\begin{equation}
\epsilon_0(\text{nn-tRVB})={\langle  \text{nn-tRVB} | \cal{H}_\text{XXZ} | \text{nn-tRVB} \rangle \over \langle  \text{nn-tRVB} | \text{nn-tRVB} \rangle }.
\label{eq:RVBener}
\end{equation}

For $J'=J$ the overlap with the exact ground state is 
$\langle \Psi_0(\text{exact}) | \text{nn-tRVB} \rangle=0.9747$, its energy is $\epsilon_0(\text{nn-tRVB})=-3.397058 J$
which is $4.07 \%$ higher than the exact ground state energy. In contrast, 
an RVB analysis of the Heisenberg antiferromagnetic model on the same cluster 
using nn singlet VBs would give a much more accurate description of the exact ground state: 
$\langle \Psi_0^\text{HAFM}(\text{exact}) | \text{nn-RVB} \rangle=0.9988$ and $\epsilon^{HAFM}_0(nn-\text{RVB})=-3.04412 J$. 
The energy of this RVB state is only 
$0.16\%$ higher than the exact ground state energy, $\epsilon^{HAFM}_0(\text{exact})= -3.052775 J $. 
Hence, it appears that the nn RVB  gives a much better description of the 
ground state of the AFM Heisenberg model than the nn-\text{tRVB} of the easy-plane XXZ magnetic exchange we consider in this work. 

We therefore consider a more general nn tRVB ansatz:
\small
\begin{eqnarray}
|\text{tRVB} \rangle=
\alpha \left( |A1B1\rangle |A2A3\rangle |B2B3\rangle  +|A1A3\rangle |A2B2\rangle |B1B3\rangle  \right.
\nonumber \\
\left. +|A1A2\rangle |A3B3\rangle |B1B2\rangle  \right)
-\beta |A1B1\rangle |A2B2\rangle |A3B3\rangle,  
\nonumber \\ 
\normalsize
%|\text{tRVB} \rangle &=& %\alpha \left( |14\rangle %|23\rangle |56\rangle  + %|13\rangle |25\rangle %|46\rangle  +|12\rangle %|36\rangle |45\rangle  \right)
%\nonumber \\
%&&-\beta |14\rangle |25\rangle %|36\rangle , 
%+ 2\gamma( (16\rangle (23\rangle (45\rangle  + (15\rangle (23\rangle (46\rangle
%\nonumber \\
%&+&(12)(34)(56)+(13)(24)(56))+2 \delta ( (15)(24)(36)
%\nonumber \\
%&+&(34)(16)(25) +(15)(24)(36)+(16)(25)(36)).
\label{eq:varRVB}
\end{eqnarray} 
%For  the anisotropic FM Heisenberg model, 
%the minimum energy of the $|t-RVB\rangle$ is achieved when:
%\begin{equation}
%\alpha=0.93788, \beta=0.33557
%\end{equation}
where $\alpha$ and $\beta$ are determined variationally. For $J'=J$ we  find an energy of 
$\epsilon_0(\text{tRVB} )=-3.453257J$ 
which is only $2.5\%$ higher than the exact ground state energy.
The overlap with the exact wavefunction is now $\langle \Psi_0(\text{exact}) | \text{tRVB} \rangle=0.98849$,
improving our previous result. For comparison a singlet-RVB version of (\ref{eq:varRVB}) for 
the Heisenberg antiferromagnet (HAF) recovers its exact ground state, {i.e.,} $\langle \text{RVB} | \Psi^\text{AFM}_0(\text{exact}) \rangle = 1$, with an energy $\epsilon_0(\text{RVB}) =\epsilon^\text{AFM}_0(\text{exact})=-5.302775 J$.

\begin{table*}
\centering
\begin{tabular}{|c |c| c c c c c c |l| l|} 
 \hline
short name & $\Gamma$ &  E & 2C$_6$ & 2C$_3$  &  C$_2$ & 3$\sigma_v$  & 3$\sigma_d$ & \text{Singlet superconducting order} & \text{Triplet superconducting order}\\  
 \hline 
$s$ & A$_1$ & 1 & 1 & 1 & 1 &  1 & 1 & \begin{tabular}{@{}l@{}} $\left[(1,1,1),(0,0,0),(1,1,1)\right]$, \\ $\left[(0,0,0),(1,1,1),(0,0,0)\right]$ \end{tabular}  & \\ 
 \hline
$i$ & A$_2$ & 1 & 1 & 1 & 1  & -1 & -1 & &  [(1,1,1),(0,0,0),(1,1,1)] \\
 \hline
$f$ & B$_1$ & 1 & -1 & 1 & -1 & 1 & -1 &  [(-1,-1,-1),(0,0,0),(1,1,1)] & [(0,0,0),(-1,-1,-1),(0,0,0)] \\
 \hline
$f'$ & B$_2$ & 1 & -1 & 1 & -1 & -1 &  1 & & [(1,1,1),(0,0,0),(-1,-1,-1)]\\
 \hline
$p$ &E$_1$ & 2 & 1 & -1  & -2 &  0  & 0
& \begin{tabular}{@{}l@{}} $\left[(1,0,-1),(0,0,0),(-1,0,1)\right]$, \\ $\left[(1,-1,0),(0,0,0),(-1,1,0)\right]$ \end{tabular} & 
\begin{tabular}{@{}l@{}} 
	$\left[(1,0,-1),(0,0,0),(-1,0,1)\right]$, \\ 
	$\left[(1,-1,0),(0,0,0),(-1,1,0)\right]$, \\ 
	$\left[(0,0,0),(2,-1,-1),(0,0,0)\right]$, \\ 
	$\left[(0,0,0),(0,1,-1),(0,0,0)\right]$ 
\end{tabular} \\
 \hline
$d$ & E$_2$ & 2 & -1 & -1   &  2 &  0 & 0
& \begin{tabular}{@{}l@{}} 
	$\left[(-1,0,1),(0,0,0),(-1,0,1)\right]$, \\ 
	$\left[(-1,1,0),(0,0,0),(-1,1,0)\right]$, \\ 
	$\left[(0,0,0),(-2,1,1),(0,0,0)\right]$, \\ 
	$\left[(0,0,0), (0,-1,1),(0,0,0)\right]$ 
\end{tabular}  &
\begin{tabular}{@{}l@{}} $\left[(1,0,-1), (0,0,0),(1,0,-1)\right]$, \\ $\left[(1,-1,0), (0,0,0),(1,-1,0)\right]$ \end{tabular}
\\
\hline 
\end{tabular}
\caption{Character table of the C$_{6v}$ with a list of the basis pairing functions allowed by symmetry in the singlet and triplet channels. The basis functions are expressed in terms of triplet/singlet pairing 
between nn sites of the lattice following the definitions of
Fig. \ref{fig:pairs} and Eq. \eqref{eq:pairs}.}
\label{table1}
\end{table*}

We extend the  variational $|\text{tRVB}\rangle$ analysis to other $J' \neq J$. 
The dependence of the overlap and the energy difference on $J'/J$ 
shown in Fig. \ref{fig:fig2}(b) indicates that the $|\text{tRVB} \rangle $ (Eq. (\ref{eq:varRVB}))
becomes a closer description of the exact ground state of the cluster as $J'/J$ 
increases since  $\langle \text{tRVB} | \Psi_0(\text{exact})\rangle \rightarrow 1$ and $\epsilon_0(\text{tRVB})-
\epsilon_0(\text{exact}) \rightarrow 0$. Furthermore, the large overlap 
found $\langle \text{tRVB} | \Psi_0(\text{exact}) \rangle \sim 1$  between the  \text{tRVB} state and the exact ground state 
in the whole $J'/J >0 $ range explored indicates that the $|\text{tRVB} \rangle$ 
is a good candidate for the ground state of the model.

The properties of quantum dimer models \cite{MoessnerRaman,contemp} are independent of whether the dimers represent singlets or triplets. This suggest that dimerised states will be common to both singlet and triplet RVB theories. Therefore, it is interesting to note that two different valence bond solids are found in the antiferromagnetic Heisenberg model on the DHL: a valance bond solid with the full symmetry of the lattice for $J'\gtrsim J$ and a valence bond solid with broken C$_3$ symmetry for  $J'\lesssim J$ \cite{Nourse22,Orus}.

%Hence, in the AF Heisenberg model, the optimized nn $|RVB\rangle$ state recovers the exact ground state energy of the model: $E_0^{exact}=-5.302775 J$.  Since the overlap is exactly 
%one, the optimized $|RVB\rangle$ state is the exact ground state of AF Heisenberg model.
%This is in contrast to the nn $| t-RVB  \rangle$ state we have constructed 
%for the anisotropic FM Heisenberg model in which  
%a nn $| t-RVB  \rangle$  is only an approximate description. 

%-Pairing symmetries of the superconducting correlations in the cluster? 

\section{Triplet RVB superconductivity}
\label{sec::TRVBSC}
Having established the tRVB state as a competitive ground state 
of the undoped model (\ref{eq:modeltJ}), we now consider the possibility of superconductivity mediated by the ferromagnetic part of the XXZ interaction. We first describe the  pairing states allowed by the symmetries of the DHL. The most favorable superconducting states are then determined by a microscopic calculation. \cite{Coleman2020,Konig2022}

\subsection{Symmetry group analysis of superconducting states}

A general phenomenological formulation of superconductivity can be achieved using Ginzburg-Landau theory in which the order parameter is obtained from general symmetry arguments. The full symmetry group of our model \eqref{eq:modeltJ} is $C_{6v} \otimes \mathcal{K} \otimes U(1) \otimes SU(2) $ where $ \mathcal{K} $ is the time-reversal symmetry group. %Here we analyze the possible superconducting states which can spontaneously break the C$_{6v}$ point group symmetry of the DHL lattice as well the global $U(1)$  symmetry broken in a superconductor. 

On doping the triplets pairs in the tRVB wavefunction become mobile leading to triplet superconductivity. Note that this condensate contains only opposite spin pairing (OSP), in contrast to the equal spin pair (ESP) found in, say, the A-phase of $^3$He \cite{Leggett1975}. Thus, one can fully describe the superconductivity by a scalar order parameter $\Delta(\bm k)$ instead of requiring the usual vector order parameter for a triplet superconductor, $\bm d(\bm k)$. However, these are trivially related via $\bm d(\bm k)=\Delta(\bm k)\bm{\hat{z}}$, where $\bm{\hat{z}}$ is the unit vector in the longitudinal direction in spin-space. Thus, all states consider also break the SU(2) symmetry associated with spin rotation. This is unsurprising as the XXZ interaction arises from spin-orbit coupling. 

We conveniently \cite{Senechal2019} express the elements of the $C_{6v}$ group in a 18-dimensional space generated by the pairing amplitudes, $\Delta_{\alpha{i\sigma},\beta{j\sigma'}}= \langle  c_{\alpha{i\sigma}} c_{\beta{j\sigma'}}\rangle$ consistent with the translational symmetry of the lattice. Defining the 18-component vector
\begin{eqnarray}
{\bf \Delta}_{\uparrow\downarrow}&=&(\Delta_{A{1\uparrow},A{2\downarrow}}, \Delta_{A{2\uparrow},A{1\downarrow}}, \Delta_{A{1\uparrow},A{3\downarrow}},
\Delta_{A{3\uparrow},A{1\downarrow}}, 
\nonumber \\
&&\Delta_{A{2\uparrow}, A{3\downarrow}}, \Delta_{A{3\uparrow}, A{2\downarrow}},
\Delta_{A{3\uparrow},B{3\downarrow}}, \Delta_{B{3\uparrow},A{3\downarrow}}, 
\nonumber \\
&&\Delta_{A{2\uparrow},B{2\downarrow}}, \Delta_{B{2\uparrow},A{2\downarrow}}, \Delta_{A{1\uparrow},B{1\downarrow}}, \Delta_{B{1\uparrow},A{1\downarrow}}, 
\nonumber \\
&&\Delta_{B{1\uparrow},B{2\downarrow}}, \Delta_{B{2\uparrow},B{1\downarrow}}, \Delta_{B{1\uparrow},B{3\downarrow}}, \Delta_{B{3\uparrow},B{1\downarrow}},  
\nonumber \\
&&\Delta_{B{2\uparrow},B{3\downarrow}}, \Delta_{B{3\uparrow},B{2\downarrow}} ),
\label{eq:pairsgroup}
\end{eqnarray}
the action of a group element in this representation, $R(a)$, on the pairing vector reads
\begin{equation}
   {\bf \Delta}^\prime_{\uparrow \downarrow}= R(a) {\bf \Delta}_{\uparrow \downarrow}.
\end{equation}

Since spin is explicitly retained in the pairing amplitudes this nomenclature allows one to consider both singlet and OSP triplet states combinations simultaneously. From the $R(a)$ obtained for all the elements, $a$, of the $C_{6v}$ character Table \ref{table1}
we can obtain the projector operators on each of its irreducible representations
\begin{equation}
P^{(\mu)}= { d_{\mu} \over g} \sum_a  \chi^{(\mu)*}(a) R(a) 
\end{equation} 
where $\chi^{(\mu)}(a)$ is the character of the $a$th element in the $\mu$th irreducible representation,  $d_{\mu}$ is the dimension of the
irreducible representation $\mu$, and $g$ is the number of symmetry elements of the group. 

\begin{figure}
   \centering
  \includegraphics[width=4.5cm]{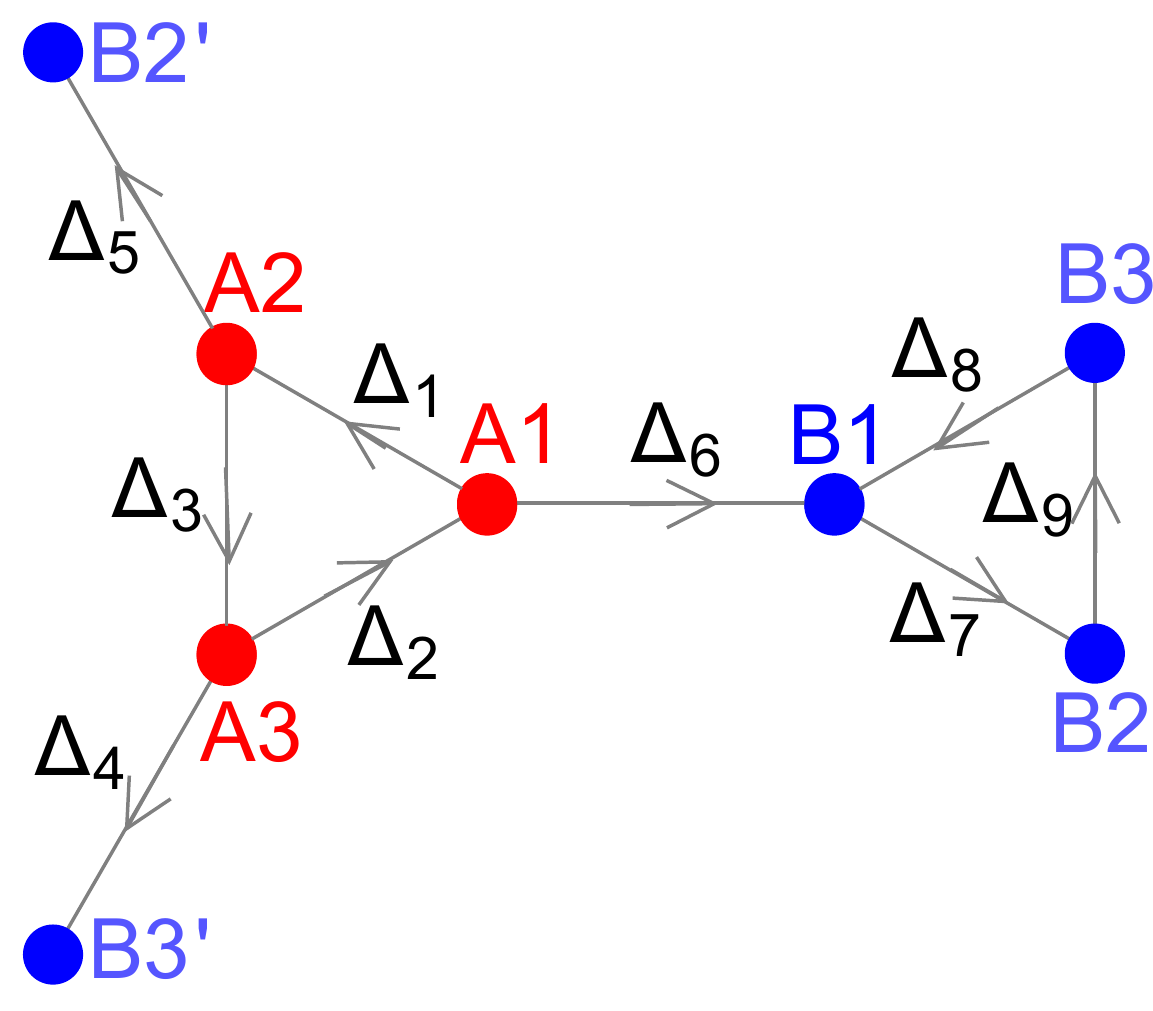}
\caption{Real space  pairing amplitudes, encoded in the $9$-dimensional 
vector, ${\bf \Delta }$, used in the mean-field analysis and linearized gap equations.  The two triangular clusters in the unit cell are shown in red (A) and blue (B). For triplet superconductivity, the pairing amplitudes against the directions indicated by the arrows  
are minus the amplitudes along the arrow directions. In the singlet case arrows should be ignored since the  pairing is non-directional. 
}
 \label{fig:pairs}
 \end{figure}

The representation of the $g=12$ elements 
of the $C_{6v}$ symmetry group in the space spanned by the 
pairing amplitudes consists of twelve $18 \times 18$ matrices. 
The eigenvectors of the projector, $P^{(\mu)}$, with eigenvalue 1 
describe basis functions of the six %$K=6$ 
irreducible representations
of the $C_{6v}$ group. These basis functions are included in Table \ref{table1}. The pairing amplitudes between the sites given in Table \ref{table1} are expressed in terms of a three-component vector of three-component vectors,
%\begin{widetext}
\begin{subequations}
\begin{equation}
{\bf \Delta}=[{\bf \Delta}_A,{\bf \Delta}_X,{\bf \Delta}_B],
\end{equation}
where,
\begin{eqnarray}
{\bf \Delta}_A &=& (\Delta_{A1,A2}, \Delta_{A3, A1}, \Delta_{A2, A3}), \\
{\bf \Delta}_X &=& (\Delta_{A3,B3}, \Delta_{A2,B2}, \Delta_{A1,B1}),  \\
{\bf \Delta}_B &=& (\Delta_{B1,B2}, \Delta_{B3,B1}, \Delta_{B2,B3}),
\end{eqnarray}
\label{eq:pairs}
\end{subequations}
where $\Delta_{\alpha i,\beta j}=(\Delta_{\alpha i\uparrow,\beta j\downarrow}+\Delta_{\alpha i\downarrow,\beta j\uparrow})/2$ for OSP triplet states and $\Delta_{\alpha i,\beta j}=(\Delta_{\alpha i\uparrow,\beta j\downarrow}-\Delta_{\alpha i\downarrow,\beta j\uparrow})/2$ for singlet pairing.
%\end{widetext}
%${\bf \Delta}$ describes 
The real space representation of  ${\bf \Delta}$ is sketched in Fig. \ref{fig:pairs}. 
%where OSP triplet  bonds, $\Delta_{\alpha i,\beta j}= ( c_{\alpha i\uparrow} c_{\beta j\downarrow} + c_{\alpha i\downarrow} c_{\beta j\uparrow})/\sqrt{2} $ are assumed. The 
The arrows in Fig. \ref{fig:pairs} describe the antisymmetry of the triplet pairs,  $\Delta_{\beta j,\alpha i}=-\Delta_{\alpha i,\beta j}$; for singlet pairing \footnote{
The $f$-wave superconducting state found in our previous paper \cite{Merino2021} was incorrectly labeled as a B$_2$ but is in fact a B$_1$ (both are $f$-wave states). The nodal structure corresponds to an $f_{x(3y^2-3x^2)}$ cubic basis function.
},  $\Delta_{\beta j,\alpha i}=\Delta_{\alpha i,\beta j}$ and the order in which the bond is
taken does not matter, so the arrows have no meaning and can be ignored. 

%-discuss the GL functional.?

%\section{Bogoliubov decoupling}

\subsection{Mean-field OSP triplet RVB theory}

With the considerations above we explore the possibility of triplet
RVB superconductivity in our model \eqref{eq:modeltJ}.
We perform a mean-field BCS decoupling of the magnetic terms
in the hamiltonian. It is convenient to write the theory in terms of triplet bond operators \cite{Momoi2009,Coleman2020},
\begin{equation}
h_{\alpha i, \beta j}^\dagger={1 \over \sqrt{2} } (c^\dagger_{\alpha i\uparrow} c^\dagger_{\beta j\downarrow}
+c^\dagger_{\alpha i\downarrow} c^\dagger_{\beta j\uparrow} ).
\end{equation}
Whence, the XXZ exchange in model \eqref{eq:modeltJ} can be expressed as
\begin{widetext}
\begin{equation}
\sum_{\langle \alpha i, \beta j \rangle} ( S^x_{\alpha i}  S^x_{\beta j} + S^y_{\alpha i}  S^y_{\beta j} - S^z_{\alpha i}  S^z_{\beta j} + {n_{\alpha i} n_{\alpha j} \over 4})= \sum_{\langle \alpha i, \beta j \rangle} h_{\alpha i ,\beta j}^\dagger h_{\alpha i, \beta j},
\end{equation}
We 
consider a BCS decoupling of the hamiltonian described through the 
pairing amplitudes:
\begin{eqnarray}
%\chi_{\alpha i, \beta j}  &=& \langle c^\dagger_{\alpha i, \sigma} c_{\beta j, \sigma} \rangle,
%\nonumber  \\
\Delta_{\alpha i, \beta j}  &=& \langle h_{\alpha i,\beta j} \rangle.
\end{eqnarray}
The mean-field hamiltonian with the renormalization effects 
due to the no double occupancy constraint, is then,
\begin{equation}
{\cal H}_{MF}={\cal H}_t+{\cal H}_J
\end{equation}
where:
\begin{eqnarray}
{\cal H}_t &=& - {\tilde t} \sum_{{\bf k}\langle  \alpha i, \alpha j \rangle,\sigma} \left( c^\dagger_{\alpha i\sigma}({\bf k}) c_{\alpha j\sigma} ({\bf k}) + h. c.  \right)-{\tilde t}' \sum_{\substack{{\bf k}, \langle A i, B i \rangle, \sigma} }\left( e^{i {\bf k}\cdot 
{\bf \delta}_{Ai,Bi} } c^\dagger_{A i\sigma}({\bf k}) c_{B i\sigma} ({\bf k}) + h.c. \right),  
\nonumber \\
{\cal H}_J =
%J \sum_{{\bf k}\langle \alpha i,  \alpha j \rangle,\sigma} \left( {\chi_{\alpha i, \alpha j} \over 2} c^\dagger_{\alpha i\sigma}({\bf k}) c_{\alpha j\sigma} ({\bf k}) + h. c.  \right)+ J'\sum_{\substack{{\bf k}, \sigma, \langle A i, B i \rangle} }\left( {\chi_{A i, B i} \over 2}  e^{i {\bf k}\cdot {\bf \delta}_{Ai,Bi}} c^\dagger_{A i\sigma} ({\bf k}) c_{B i\sigma}({\bf k})  + h.c. \right)  
%\nonumber \\
&-&{ {\tilde J} \over \sqrt{2}} \sum_{{\bf k}\langle \alpha i, \alpha j \rangle} \left( \Delta_{\alpha i,\alpha j}  c^\dagger_{\alpha i\uparrow}({\bf k}) c^\dagger_{\alpha j\downarrow} (-{\bf k}) - \Delta_{\alpha i,\alpha j}  c^\dagger_{\alpha  j\uparrow}({\bf k}) c^\dagger_{\alpha i\downarrow} (-{\bf k})  \right ) +h.c.
\nonumber \\
&-& { {\tilde J}' \over \sqrt{2} } \sum_{{\bf k}\langle A i, B i \rangle} \left(  \Delta_{A i,B i} e^{i {\bf k} \cdot {\bf \delta}_{Ai,Bi}}  c^\dagger_{A i\uparrow}({\bf k}) c^\dagger_{B i\downarrow} (-{\bf k}) -  \Delta_{A i,B i} e^{-i {\bf k} \cdot {\bf \delta} _{Ai,Bi}} c^\dagger_{B i\uparrow}({\bf k}) c^\dagger_{A i \downarrow} (-{\bf k})  \right) + h.c.
\nonumber \\
&+& {\tilde J}  \sum_{\langle \alpha i, \alpha j \rangle } |\Delta_{\alpha i,\alpha j}|^2
%- | \chi_{\alpha i, \alpha j}|^2 \right) 
+ {\tilde J}' \sum_{\langle A i, B i \rangle} |\Delta_{A i,B i}|^2 
%- |\chi_{A i, B i}|^2  \right)
+\mu \sum_{\alpha{\bf k}\sigma} c^\dagger_{\alpha\sigma} ( {\bf k}) c_{\alpha\sigma} ({\bf k}),
\label{eq:hamMF}
\end{eqnarray}
\end{widetext}
where we have made the Gutzwiller approximation (GA), which yields renormalized parameters: ${\tilde t}/t = {\tilde t}'/t'= {2 \delta / (1 + \delta)} \equiv g_t$ and ${\tilde J}/J={\tilde J}'/J'= {4 / (1 + \delta )^2} \equiv g_J$.
%with $i,j=1,2,3$ enumerating the sites inside each triangular unit and $\alpha=A, B$ referring to any two nearest-neighbor triangular units of the lattice. Thus the sums are restricted
%to nn sites in the same triangle, $\langle \alpha i , \alpha j \rangle$
% or nn sites between two neighbor triangles, $\langle A i, B i \rangle $. 
The numbering of sites in two neighboring $A, B$ triangles together with the different pairing amplitudes entering the Gorkov decoupling are shown in Fig. \ref{fig:pairs}.

\begin{figure}
\centering
    \includegraphics[width=8cm,clip=]{ 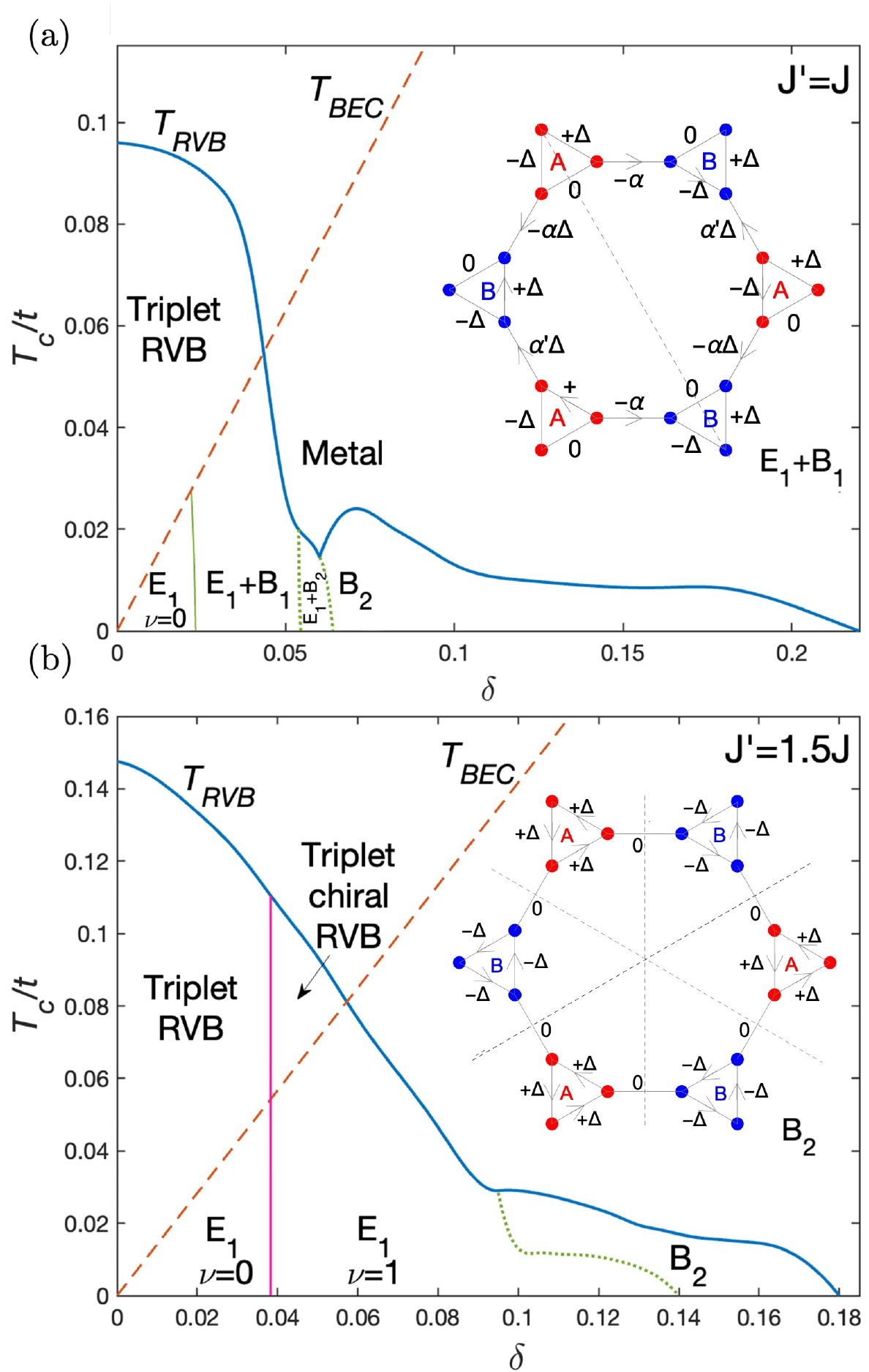} 
   \caption{ %Triplet \text{RVB} superconduc5tivity in the $t$-$J$ model on the decorated honeycomb lattice. 
   $T$ vs $\delta$ phase diagrams of the $t$-$J$ model \eqref{eq:modeltJ} from triplet-RVB theory for %(\ref{eq:modeltJ}) 
   (a) $J'/J=1$ and (b) $J'/J=1.5$.
   Superconductivity occurs when $T<T_{\text{BEC}}$ (red dashed line)  and $T< T_{\text{RVB}}$ (blue line). 
   While the transition at $T_{\text{RVB}}$ (blue solid line) is second order, the transitions between different superconducting orders can be first (green dotted lines) or second (green solid lines) order. A second order topological transition (pink solid line) separates E$_1$ ($\nu=0$) and E$_1$
   ($\nu=1$) in (b).  The insets in (a) and (b) illustrate the pairing patterns of the $p+f$ (E$_1$+B$_1$) and $f$ (B$_2$) solutions including their nodal lines in real space
   (black dashed lines). 
%At high dopings the pairing $SC(B_2)$ state is favored at (a) $J'/J=1$ and (b) $J'/J=1.5$. This triplet superconducting pattern (illustrated in the inset) displays three nodal lines being consistent with an $f$-wave triplet superconductor. (b) At low dopings and $J'=1.5J$ such pairing state becomes a topological $E_1$ superconductor ($TSC(E_1)$) featured by a Chern number of $\nu=1$ at low dopings.
%For temperatures between $T_{BEC}<T<T_{RVB}$ a topologically non-trivial triplet RVB state arises. 
%(a) Temperature vs. doping phase diagram of the model (\ref{eq:model})
%showing the triplet-RVB phase and superconducting phases.  (b) The triplet superconducting 
 %pattern $SC(E_1)$ at low hole dopings. (c) The triplet pairing $SC(B_2)$ favored at higher dopings.
 %This pairing state has a single nodal line (dashed line) which is consistent with a $p_x-p_y$ triplet superconductor. 
 We have used $J=0.1t$ and $J'/J=t'^2/t^2$.
}
 \label{fig:phased}
 \end{figure} 
After diagonalization, the hamiltonian reads
\begin{eqnarray}
H_{MF} &=& \sum_{m,{\bf k}, \sigma} \omega_m({\bf k} ) \left( \gamma^\dagger_{m \sigma} ({\bf k}) \gamma_{m \sigma} ({\bf k}) - {1 \over 2}  \right) 
\nonumber \\
&+& {\tilde J} \sum_{\langle \alpha i, \alpha j \rangle }  |\Delta_{\alpha i,\alpha j}|^2 
%-|\chi_{\alpha i, \alpha j}|^2) 
+ {\tilde J}' \sum_{\substack{\langle A i, B i \rangle} } |\Delta_{A i,B i}|^2 
%-|\chi_{A i,B i}|^2),
\nonumber \\
\end{eqnarray}
where $\omega_m({\bf k})>0$ with $m=1,...,6$ are the positive Bogoliubov quasiparticle dispersions.

The free energy of the system is
\begin{eqnarray}
\Phi &=& -{1 \over \beta} \sum_{m,{\bf k}, \sigma} \ln(1+ e^{-\beta \omega_m({\bf k})} )- \sum_{m,{\bf k}} \omega_m({\bf k}) + 6 N_s \mu 
\nonumber \\
&+& {\tilde J} \sum_{\langle \alpha i, \alpha j \rangle } \left(  |\Delta_{\alpha i,\alpha j}|^2 \right)
%-|\chi_{\alpha i, \alpha j}|^2 \right)
+ {\tilde J}' \sum_{\substack{\langle A i, B i \rangle} } \left( |\Delta_{A i,B i}|^2  \right)
%-|\chi_{A i, B i}|^2   \right).
\end{eqnarray}
From the minimization of the free energy we arrive at a set of self-consistent equations (SCEs):
\begin{eqnarray}
%\chi_{\alpha i,\alpha j} &=& {1 \over {\tilde J} N_s} \sum_{m,{\bf k},\sigma} \left( f(\omega_m ({\bf k}) ) -{1 \over 2} \right) \left( {\partial \omega_m({\bf k}) \over \partial \chi^*_{\alpha i, \alpha j} }\right),
%\nonumber \\
%\chi_{A i, B i} &=& {1 \over {\tilde J}' N_s} \sum_{m,{\bf k},\sigma} \left( f(\omega_m ({\bf k}) ) -{1 \over 2} \right) \left( {\partial \omega_m({\bf k}) \over \partial \chi^*_{A i, B i } } \right),
%\nonumber \\
\Delta_{\alpha i,\alpha j} &=& -{1 \over {\tilde J} N_s} \sum_{m,{\bf k},\sigma} \left( f(\omega_m ({\bf k}) ) -{1 \over 2} \right) \left( {\partial \omega_m({\bf k}) \over \partial \Delta^*_{\alpha i, \alpha j} }\right),
\nonumber \\
\Delta_{A i,B i} &=& -{1 \over {\tilde J}' N_s} \sum_{m,{\bf k},\sigma} \left( f(\omega_m ({\bf k}) ) -{1 \over 2} \right) \left( {\partial \omega_m({\bf k}) \over \partial \Delta^*_{A i, B i } }\right),
\nonumber \\
\delta &=& -{1 \over 6 N_s} \sum_{m,{\bf k}, \sigma} \left( f(\omega_m({\bf k}))- {1 \over 2} \right) \left( {\partial \omega_m({\bf k})\over \partial \mu} \right),\notag\\
\label{eq:sce}
\end{eqnarray}
which we solve numerically. %The doping is defined as $\delta=1-n$, where $n$ is the average electron occupancy per site. 
%Note that the final GA projected solution of the $t$-$J$ model effectively implies substituting $(J, J') \rightarrow ({\tilde J}, {\tilde J'})=(g_J J, g_{J'} J')$ and $(t, t') \rightarrow ( {\tilde t}, {\tilde t'})=(g_t t, g_{t'} t')$ with $g_t=g_{t'}= {2 \delta \over 1 + \delta}$ and $g_{J}=g_{J'}= {4 \over (1 + \delta )^2}$,in the free energy and the above set of self-consistent equations.

\subsection{Phase diagrams and spinon dispersions} 

We now consider { uniform} metallic and superconducting solutions of the SCEs (\ref{eq:sce}). As in singlet RVB theory, there are two temperature scales: $T_{\text{RVB}}$ and $T_{\text{BEC}}$, which signify the onset of fermionic pairing and coherence respectively (the latter is corresponds to Bose-Einstein condensation in the slave boson reformulation of theory \cite{Kotliar-Liu}). 

%At 
%$\delta=0$ the system is an insulating RVB spin liquid for $T < T_{\text{RVB}}$ and $T_{\text{BEC}}=0$. 
%The short-range spin correlations of the RVB state survive under small hole doping leading
%to an unconventional metal with a spin gap for $T_{\text{RVB}}>T>T_{\text{BEC}}>0$. 
%For large doping in the regime $T_{\text{BEC}}>T>T_{\text{RVB}}>0$ there is coherence without pairing and the system is a Fermi liquid.
%Near optimal doping, when $T>T_{\text{BEC}}, T_{\text{RVB}}$ we have a strange metal \cite{Hussey}.

%For $T<T_{\text{BEC}}, T_{\text{RVB}}$,
%the added holes pair coherently and the system becomes a high-$T_c$ superconductor. 
%On the square lattice, relevant to the cuprate 
%CuO$_2$ planes, $d_{x^2-y^2}$ RVB superconductivity is 
%found \cite{Anderson1987}. 
%On the triangular lattice relevant to organic superconductors and spin liquids and Na$_x$CoO$_2$ $d_{x^2-y^2}+id_{xy}$ RVB superconductivity is 
%found \cite{Powell2007}.
%Our recent application of singlet RVB theory \cite{Merino2021}
%to the isotropic $t$-$J$ model on the DHL has shown that various superconducting states 
%allowed by symmetry (see Table \ref{table1}) are possible and that many of these occur in the flat band. Most exotically, we find f-wave singlet pairing. 

tRVB theory allows us to characterize the 
superconducting states emerging in the  $t$-$J$ model
for different $J'/J$ and $\delta$. The phase diagrams of the 
$J'=J$ and $J' =1.5 J$  are shown in Fig. \ref{fig:phased}. 
The phase diagrams share similar qualitative features. 
At zero doping there is a ferromagnetic quantum spin liquid \cite{Khosla2017}.
For $T< T_{\text{RVB}}, T_{\text{BEC}}$ there is pairing and coherence, so we have triplet superconducting phases.
For $T_{\text{RVB}}<T<T_{\text{BEC}}$ there is coherence but no pairing, and we have a conventional Fermi liquid. 
For $T_{\text{RVB}}, T_{\text{BEC}} < T$ there is neither pairing nor coherence, so we have a strange metal.
All of these phases are closely analogous to the phases in the same regimes of the singlet RVB theory \cite{Lee2006}.

However, for  $T_{\text{BEC}}<T<T_{\text{RVB}}$, there is pairing but no coherence. In the singlet RVB theory this is interpreted as a spin gap or pseudogap \cite{Lee2006}. In the tRVB theory we need to take care as we are dealing with triplet, rather than singlet, pairs. Furthermore, the strong SMOC assumed in the derivation of the XXZ model (Eq. (\ref{eq:xxz})) implies that the `spins' in our model are actually spin-orbit entangled pseudospins. Therefore, we expect a gap for longitudinal magnetic fields, but no gap for transverse magnetic fields in the tRVB pseudogap phase. This contrasts with the singlet-RVB  state where the (pseudo)gap  is apparent  
for any magnetic field direction.    
For the tRVB state for sufficiently large longitudinal fields (relative to the strength of the spin-orbit coupling) we expect, on  general energetic grounds, a transition where the triplet order parameter rotates to be perpendicular to the field \cite{Ben1D}, this is analogous Fr\'eedericksz transition in liquid crystals and $^3$He in slab geometries \cite{Vollhardt}.

In both cases at the larger dopings at which weak coupling theory is 
relevant the most
stable superconducting solution has $f$-wave (B$_2$) symmetry, 
\begin{equation}
{\bf \Delta}(\text{B}_2)=[(1,1,1),(0,0,0), (-1,-1,-1) ],
\end{equation}
which is sketched in the inset of Fig. \ref{fig:phased}(b). As expected, given the scalar order parameter, this $f$-wave triplet superconducting 
solution displays three nodal lines in real space. 

In the strong coupling regime, $\delta \rightarrow 0$, a $p+ip$ (E$_1$) superconducting phase is found to be the ground state for both exchange  $J'/J=1$ and $J'/J=1.5$. This state displays a complex order parameter given by: 
\begin{eqnarray}
	 {\bf \Delta}(\text{E}_1)
	=&&
	%[e^{i \alpha_1} (1,-1,0)+ (1,0,-1),
	%\nonumber \\
	%&& \Delta e^{i \alpha_2}  (e^{i 2 \pi/3}, e^{-i 2 \pi/3}, 1),
	%\nonumber \\
	%&& e^{i \alpha_1} (-1,1,0)+ (-1,0,1) ],
	% \nonumber\\
	[ (1,0,-1),(0,0,0), (-1,0,1) ]
	\nonumber\\
	&&+
	e^{i\alpha_1} [ (1,-1,0),(0,0,0), (-1,1,0) ] 
	\nonumber\\
	&&-\Delta' \big( e^{i\alpha_2}[ (0,0,0),(2,-1,-1), (0,0,0) ]
	\nonumber\\
	&&+\sqrt{3} e^{i\alpha_1}[ (0,0,0),(0,1,-1), (0,0,0) ] \big),
%	 =[e^{i (\alpha_1 - \alpha_2)} (1,-1,0)+ e^{-i \alpha_2}(1,0,-1),
%	 \nonumber \\
%	 \Delta   (e^{i 2 \pi/3}, e^{-i 2 \pi/3}, 1),
%	 \nonumber \\
%	 e^{i (\alpha_1 - \alpha_2)} (-1,1,0)+ e^{-i \alpha_2}(-1,0,1) ],
%	 \\
%	 =[ (e^{i (\alpha_1 - \alpha_2)},-e^{i (\alpha_1 - \alpha_2)},0)+ (e^{-i \alpha_2},0,-e^{-i \alpha_2}),
%	 \nonumber \\
%	 \Delta   (e^{i 2 \pi/3}, e^{-i 2 \pi/3}, 1),
%	 \nonumber \\
%	  (-e^{i (\alpha_1 - \alpha_2)}, e^{i (\alpha_1 - \alpha_2)}, 0)+ (-e^{-i \alpha_2},0,e^{-i \alpha_2}) ], \nonumber
%	  \\
%	  &=&\big[ (e^{i (\alpha_1 - \alpha_2)} + e^{-i \alpha_2},-e^{i (\alpha_1 - \alpha_2)},-e^{-i \alpha_2})
%	  \nonumber \\
%	  &&\Delta   (e^{i 2 \pi/3}, e^{-i 2 \pi/3}, 1),
%	  \nonumber \\
%	  &&(-e^{i (\alpha_1 - \alpha_2)} -e^{-i \alpha_2}, e^{i (\alpha_1 - \alpha_2)}, e^{-i \alpha_2}) \big]
%	  \\
%	  &=&\big[ (e^{-i \beta_1} + e^{-i \alpha_2},-e^{-i \beta_1},-e^{-i \alpha_2})
%	  \nonumber \\
%	  &&\Delta   (e^{i 2 \pi/3}, e^{-i 2 \pi/3}, 1),
%	  \nonumber \\
%	  &&(-e^{-i \beta_1} -e^{-i \alpha_2}, e^{-i \beta_1}, e^{-i \alpha_2}) \big],
	\label{eq::E1TRS}
\end{eqnarray}
with lightly $\delta$-dependent phases around $\alpha_1 = \pi/3$ and $\alpha_2 = \pi/6$ (up to numerical accuracy) %\textcolor{red}{How close are these? Should we just write "=" and write "up to numerical accuracy" or something similar}
and the amplitude $\Delta'$ strongly dependent on the doping.
%\textcolor{red}{$(\alpha_2 - \alpha_1)=\beta_1\sim \pi/2 - \pi/3= \pi/6$} 
Clearly, this state breaks time reversal symmetry.

However, important differences between the two phase diagrams appear in the intermediate doping range.
For $J'=J$ the E$_1$ and B$_2$ pairing states are separated by two superconducting states obtained from the mixing of two different irreducible representations of Table \ref{table1}. 
At $\delta >0.023$, a second order transition to a $p+f$ (E$_1$+B$_1$) symmetry superconducting state occurs, where
\begin{equation}
{\bf \Delta}(\text{E}_1+\text{B}_1)=[(1,0,-1),(-\alpha,\alpha',-\alpha), (-1,0,1) ],
%\end{equation}
\label{E1B1}
\end{equation}
has the lowest energy. The intertriangle amplitudes ratio $\alpha'/\alpha$ has a doping dependence which is analyzed in  Appendix \ref{app:hybridsc} (see especially Fig. \ref{fig::aphaE1B1}). By raising the doping above $\delta\sim 0.053$ a first order transition to a $p+f'$ (E$_1$+B$_2$) pairing state occurs, which order parameter reads
\begin{equation}
{\bf \Delta}(\text{E}_1+\text{B}_2)=[(1,1,\eta),(-\xi,\xi,0), (-1,-1,\eta) ],
%\end{equation}
\end{equation}
with $0<\eta,\xi<1$. Then at $\delta=0.064$ the system transits to the pure f-wave (B$_2$) described above. 
These pairing states come from linear combinations of $p$ (E$_1$) and $f$ (either  B$_1$ or B$_2$) superconducting solutions (see Table. \ref{table1}).
%, which become degenerate as the effective coupling is increased (see discussion of $T_c$s in the following subsection). 
  \begin{figure}[t!]
\centering
    \includegraphics[width=8.5cm,clip=]{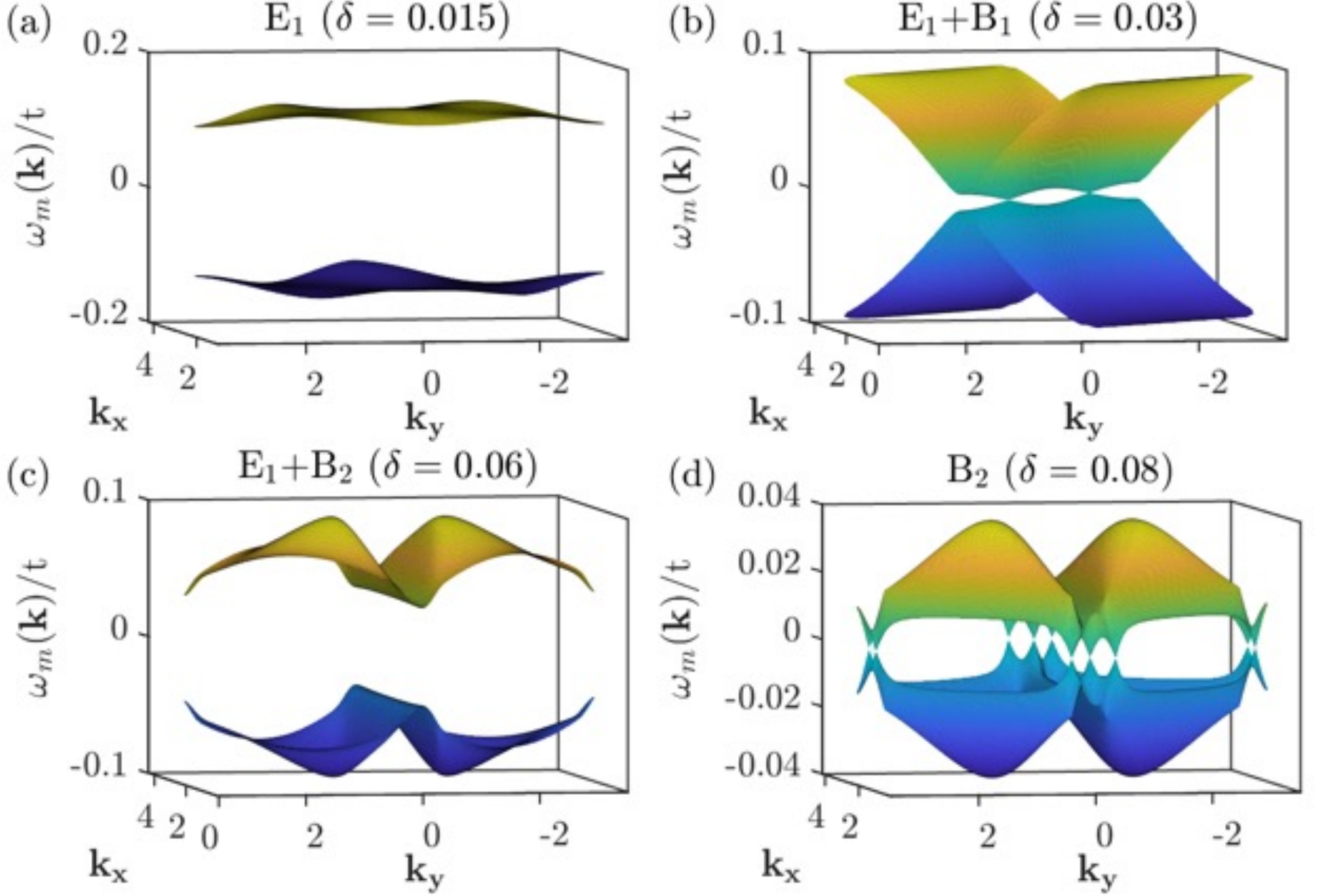} 
   \caption{Mean-field BdG bands at a fixed temperature $T=0.01t$ for the different pairing symmetries at $J'/J=1$. Only the occupied, $-\omega_{1}(\bf{k})$, and unoccupied, $\omega_1(\bf{k})$, %\textcolor{red}{I don't think this notation is concistent with the main text - I think we write $\omega_m$ in Eq 25 with m=1...9, I guess $\omega+{-1}$ should really be $-\omega_1$, but I didn't want to change this without talking to you. Please fix - or just delete the $\omega$s as they are not really needed.}, 
   bands closest to the Fermi level are shown for simplicity.}
 \label{fig:BdGbands}
 \end{figure}
These `hybrid' superconducting states are forbidden in the linearized gap equations (see discussion of $T_c$s in the following subsection), as every solution must transform according to a single representation of C$_{6v}$  \cite{Sigrist1991}. However, when the superconducting state lowers the symmetry of the system to ${\cal G}$, a subgroup of C$_{6v}$, then any harmonic transforming according to the same irreducible representation of ${\cal G}$ as the superconducting order parameter can  mix without breaking additional symmetries and hence causing a phase transition (immediately) below $T_c$. A full mathematical discussion of this is given in Appendix \ref{app:hybridsc} based on the the general framework  \cite{Sigrist1991}. For $p$-wave (E$_1$) solutions on the DHL the only other representations of C$_{6v}$ that can mix are $f$-wave (B$_1$ and B$_2$). Thus, in  $p+f^{(\prime)}$ states are common on the DHL as they often have lower energy than pure $p$-wave states.

%\begin{equation}
%{\bf \Delta}(E_1+B_1)=((1,0,-1),(-\alpha,\alpha,-\alpha), (-1,0,1) ).
%\end{equation}
%For $J'=1.5J$ the range of doping where the f-wave (B$_2$) state dominates is reduced and a first order transition from $f$-wave ($B_2$) to $p+ip$-wave ($E_1$) occurs at $\delta \approx 0.14$ with complex order parameter given by: 
For $J'=1.5J$ the complex $p+ip$ solution \eqref{eq::E1TRS} survives for much higher dopings relegating the B$_2$ state to a small doping region. The E$_1$ superconducting parameter  increases rapidly as doping decreases ($||\bm\Delta(\text{E}_1)||\gg 1$ for $\delta\rightarrow 0$).
Furthermore, we find a second order transition at $\delta_c=0.039$ above which the $p+ip$ superconducting state becomes topologically non-trivial with a Chern number $\nu=1$.
The $p+ip$ (E$_1$) topological superconductor, will be  discussed in detail in Sec. \ref{sec::TSC}.

The superconducting order parameter in reciprocal space can be quite different to its real space counterpart as expected in a multiband system.\cite{Senechal2019}  In Fig. \ref{fig:BdGbands} we show the doping dependence of the lowest energy BdG excitations for $J'=J$. In marked contrast to one band systems, the nodal structure of the dispersions of the present multiband system (with many atoms per unit cell), is very different to the one in real space. For instance, the $f$-wave (B$_2$) solution of Fig. \ref{fig:BdGbands}(d) displays only isolated nodal points in contrast to the three nodal lines occurring in real space.
%While in single band superconductors the real space nodal structure
%should coincide with the one in momentum space since they are related by a simple
%Fourier transform, this is not the case in multiband superconductors.\cite{Senechal2019} \textcolor{red}{Is there a nice reference for this - how well is it understood?} \textcolor{blue}{Have a look at the reference I added. I think it is understood that, in general, the nodal structure of a multiband system in reciprocal space is non-universal.}
%Similarly, the $f$-wave structure of the B$_2$ solution in real
%space becomes a \textcolor{blue}{nodal line}  in momentum space. 
Similarly, the $p+f$ (E$_1$+B$_1$) state shown in \ref{fig:BdGbands}(b) is characterized by two nodal points along $k_y=k_x$ in the BdG bands
%is ungapped \textcolor{red}{not sure what this means - I don't think you mean gapless (no gap anywhere) - seems to have a node in Fig 4} \textcolor{green}{we mean there are touching points so it is ungapped instead of gapped.}, 
although the superconducting state has a single nodal line in real space (see inset of Fig. \ref{fig:phased}(a)). The $p+f$ (E$_1$+B$_2$) state is gapped over the entire Brillouin zone in contrast to the three nodal lines occurring in real space.
%Talk about E_1 at J'=J and says it remains the dominating one a huge range of dopings at J'=1.5J.
Only in the $p+ip$ (E$_1$) superconducting phases the order parameter is found to be fully gapped consistently in real and reciprocal space as illustrated in Fig. \ref{fig:BdGbands}(a) for $\delta=0.015$. At the critical lines of the Fig. \ref{fig:phased} phase diagrams separating E$_1$ solutions with different topological properties nodal points occur.

%For $J'=1.5J$ the $p+ip$ (E$_1$) superconductor is the ground state for the low doping region  (Fig. \ref{fig:phased}). As previously mentioned around $\delta_c\sim 0.039$ the $p+ip$ superconductor turns topological with a Chern number $\nu=1$. This topological transition is illustrated in Fig. \ref{fig:top_trans} of the later section Sec. \ref{sec::TSC} where Bogoliuvob dispersions around $\delta_c$ are shown.
%Thus, in contrast to the $J'=J$ case, the dispersion of the Bogoliubov excitations for $J'/J=1.5$ are fully gapped up to intermediate-large dopings ($\delta=0.14$).
%%For large dopings the occupied bulk BdG bands have 
%%$\nu=1$ characterizing the E$_1$ topological superconductor, TSC(E$_1$).  
%%$\delta<0.039$ gap is closed 
%\textcolor{red}{this seems inconsistent with the figure, which is clearly gapped on the left hand side}\textcolor{green}{You're right, this is way we refer to Fig 11 in sec V where gap is clearly closed and reopened. } 
%and the TSC becomes topologically trivial $\delta< 0.039$, SC(E$_1$) (see Fig. \ref{fig:top_trans} in the Sec. \ref{sec::TSC} for details). 
\begin{figure}
   \centering
  \includegraphics[width=7.5cm]{ 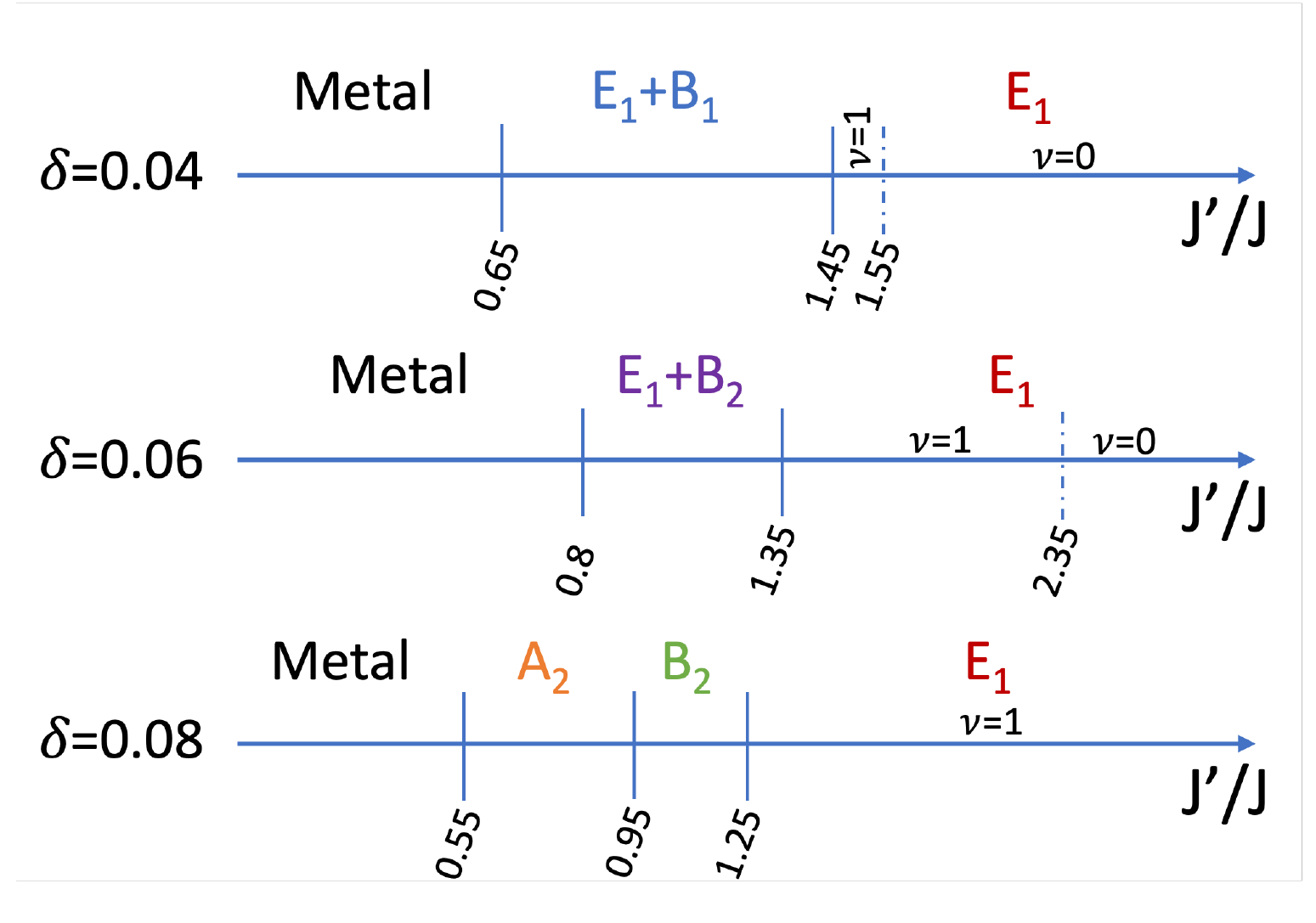}
\caption{Dependence of superconducting states on dimerization/trimerization, $J'/J$. The phase diagrams of model (\ref{eq:modeltJ}) at three different dopings: $\delta=0.04,0.06,0.08$ are shown at fixed $T=0.01$.
We have used $J=0.1t$ and $J'/J=t'^2/t^2$.
}
 \label{fig:TSCJ}
 \end{figure} 
Thus an overall trend emerges:  at strong coupling, $\delta \rightarrow 0$,
when the system is dominated by magnetic exchange the SC 
is characterized by either having isolated nodal points or  
being fully gapped. In contrast,  the weak coupling solutions
at large $\delta$ are dominated by the kinetic energy and the gap has nodal points in reciprocal space. Hence, gapless
superconductivity is more common in the weak coupling regime than at strong coupling.
%\textcolor{red}{I don't understand how this paragraph follows from the analysis above.}\textcolor{blue}{This is what we can conclude from 
%direct observation of our results: nodal lines are more common in the weak coupling limit.}
 
%\textcolor{red}{In  Fig. \ref{fig:TSCJ} we explore the evolution of the superconducting order as $J'/J$ is varied  at  fixed doping, $\delta=0.04$, for three different temperatures.
%The reason why the TSC(E$_1$) becomes favorable at large $J'/J$ is 
%that the anisotropic FM exchange lattice of the DHL
%resembles that of a honeycomb lattice in the $J'/J \rightarrow \infty$ limit. The $t-J$ model with anisotropic FM exchange on a honeycomb lattice should host chiral $p_x+i p_y$ superconductivity under hole doping as theoretically found in triangular and square lattices. \cite{Konig2022} 
%A transition from the $p+f$ (E$_1$+B$_1$) superconductivity to the TSC (E$_1$) occurs at a critical value
%$J'/J \sim 1.45 $ independent of temperature. %(*we should include more temperatures to see when does the 
%$TSC(E_1)$ state goes away?! *).
%While at $T=0$ there is a single $p+f$ (E$_1$+B$_1$) to TSC (E$_1$) transition, a second transition back to the $p+f$ (E$_1$+B$_1$) phase occurs at finite $T$ for large $J'/J$. The region of stability of the TSC (E$_1$) phase shrinks with increasing temperature until it disappears completely around $T\sim 0.11t$ (not shown). The $T=0$ $p+f$ (E$_1$+B$_1$) phase for $J'\ll J$ is very fragile since a small $T$ increase induces a metallic state.}
Hybrid and topological superconductivity are not peculiar to the specific $J'/J$ explored in Fig. \ref{fig:phased}.
The phase diagrams of Fig. \ref{fig:TSCJ} show that  the hybrid E$_1$+B$_1$ and E$_1$+B$_2$ states occur at low $\delta \sim 0.04-0.06$ in an extend range of $J'/J$, consistent with the analysis of Appendix \ref{app:hybridsc}.
At larger dopings, say $\delta=0.08$, pure A$_2$, B$_2$ and E$_1$ are favored, respectively, as $J'/J$ increases consistent with the 
weak coupling analysis of Appendix \ref{app::SCTc}. On the other hand, topological 
$p+ip$ (E$_1$) superconductivity (with $\nu=1$) is favored in the dimerized regime, $J'/J>1$,  and increasing $\delta$, Fig. \ref{fig:TSCJ}. 
As doping increases the topological transition to the trivial E$_1$ superconducting state ($\nu=0$) occurs deeper into the dimerized limit. Hence, hybrid and topological superconductivity are robust solutions of the SCE of model (\ref{eq:modeltJ}).

%\textcolor{red}{Maybe move everything above here into one of the earlier sections as it's not really about the topological nature of the SC} \textcolor{blue}{Ok, done}

\subsection{ Critical temperatures and superconducting pairing states}
\label{sec:SCpairing}
\begin{figure}
	\centering
	\includegraphics[width=8.6cm]{ 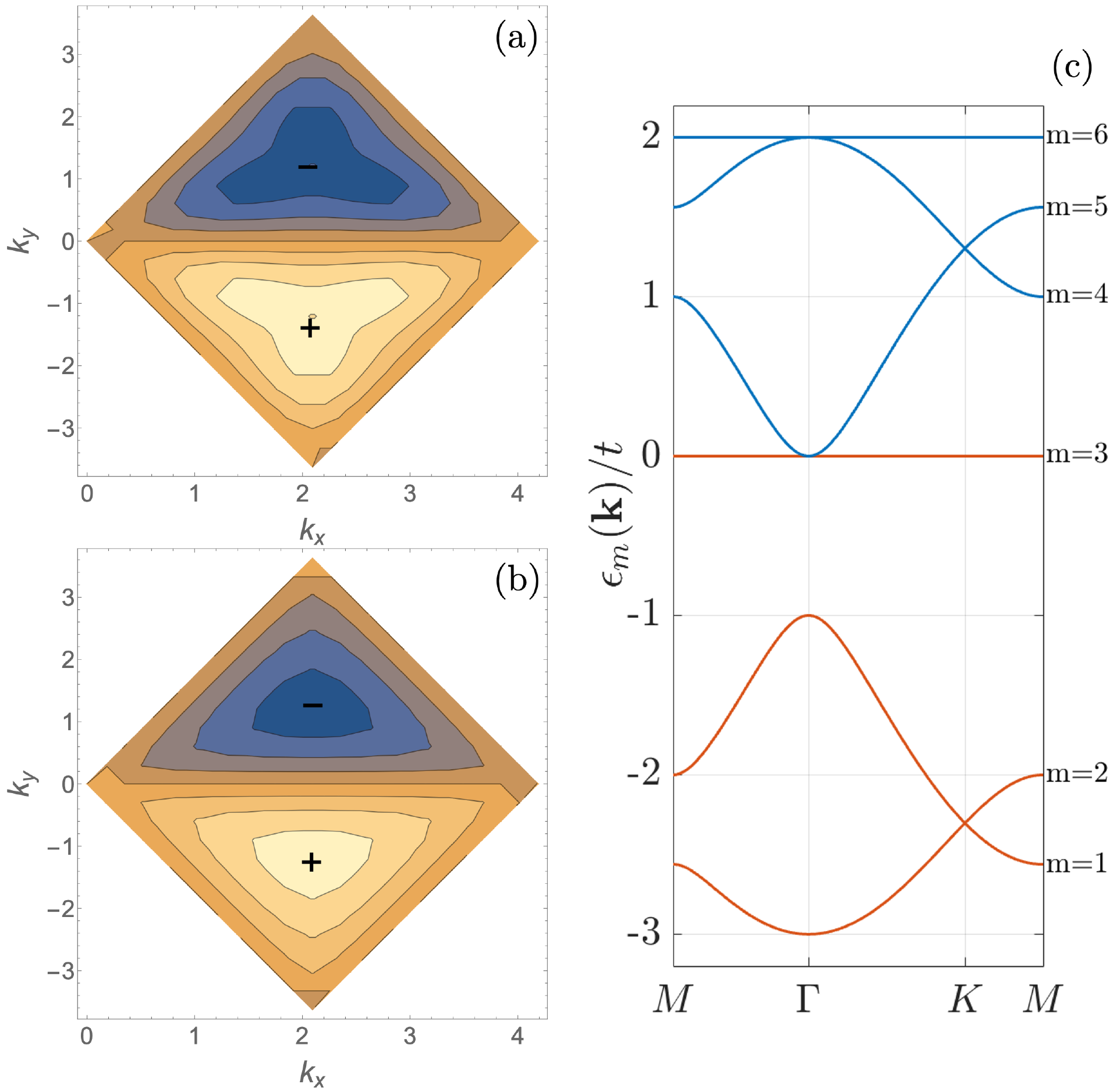}
	%%%includegraphics[width=4.25cm]{trvbgap22.pdf}
	%%%includegraphics[width=4.25cm]{trvbgap33.pdf}
	\caption{Triplet gap function projected onto the occupied bands closest to Fermi energy.   The intraband gaps, $\Delta_{22}({\bf k})$ (a)
		and $\Delta_{33}({\bf k})$ (b) 
		corresponding to the $f$-wave (B$_2$) solution for $\delta=0.09$ and $J'=J$ are shown. 
		Note the different scale of the pairing amplitude
		in the two figures:  (a) -0.068-0.068;   (b) -0.64-0.64. (c) Bare bands (eigenstates of $\mathcal{H}_t$, Eq. \eqref{eq:hamMF} for $\tilde{t}=t=\tilde{t}'=t'$) .
	}
	\label{fig:gap}
\end{figure} 
Additional understanding can be gained from a weak coupling microscopic analysis 
to obtain the $T_c$s of the different possible pairing solutions. 
The set of SCE (\ref{eq:sce}) can be linearized by expanding the quasiparticle energies
around $T_c$. This yields
\begin{equation}
 \omega_m({\bf k})=\epsilon_m({\bf k})+\sum_{n} { |\Delta_{mn}({\bf k})|^2 \over \epsilon_m({\bf k}) + \epsilon_n({\bf k})},
 \label{eq:omegam}
 \end{equation}
where the $\epsilon_m({\bf k})$ are the eigenvalues of the kinetic energy part of the 
mean-field part of the hamiltonian, ${\cal H}_t$,  Eq. (\ref{eq:hamMF}), {with $m$ running from the lowest to highest energy band  (shown in Fig. \ref{fig:gap}(c))}. The gap functions projected onto the different bands, $\Delta_{mn}({\bf k})$, read:
\begin{equation}
\Delta_{mn}({\bf k})=\sum_p {\tilde J_p} \Omega^p_{mn}({\bf k}) \Delta_p, 
\label{eq:gap}
\end{equation}
where $m,n$ denote the different bands and $p=1,..,9$ with $\Delta_p$, the real space Cooper pairing amplitudes of Eq. (\ref{eq:pairs}), shown in 
Fig. \ref{fig:pairs}, and ${\tilde J_p}={\tilde J}$ for $p=1,2,3$ and $p=7,8,9$
 while 
${\tilde J_p}= {\tilde J'}$ for $p=4,5,6$, and the $\Omega^p_{mn}({\bf k})$ are products of Bloch wavefunction coefficients  
defined in Eq. (\ref{eq:Omega}).
\begin{figure}
	\centering
	\includegraphics[width=8cm]{ 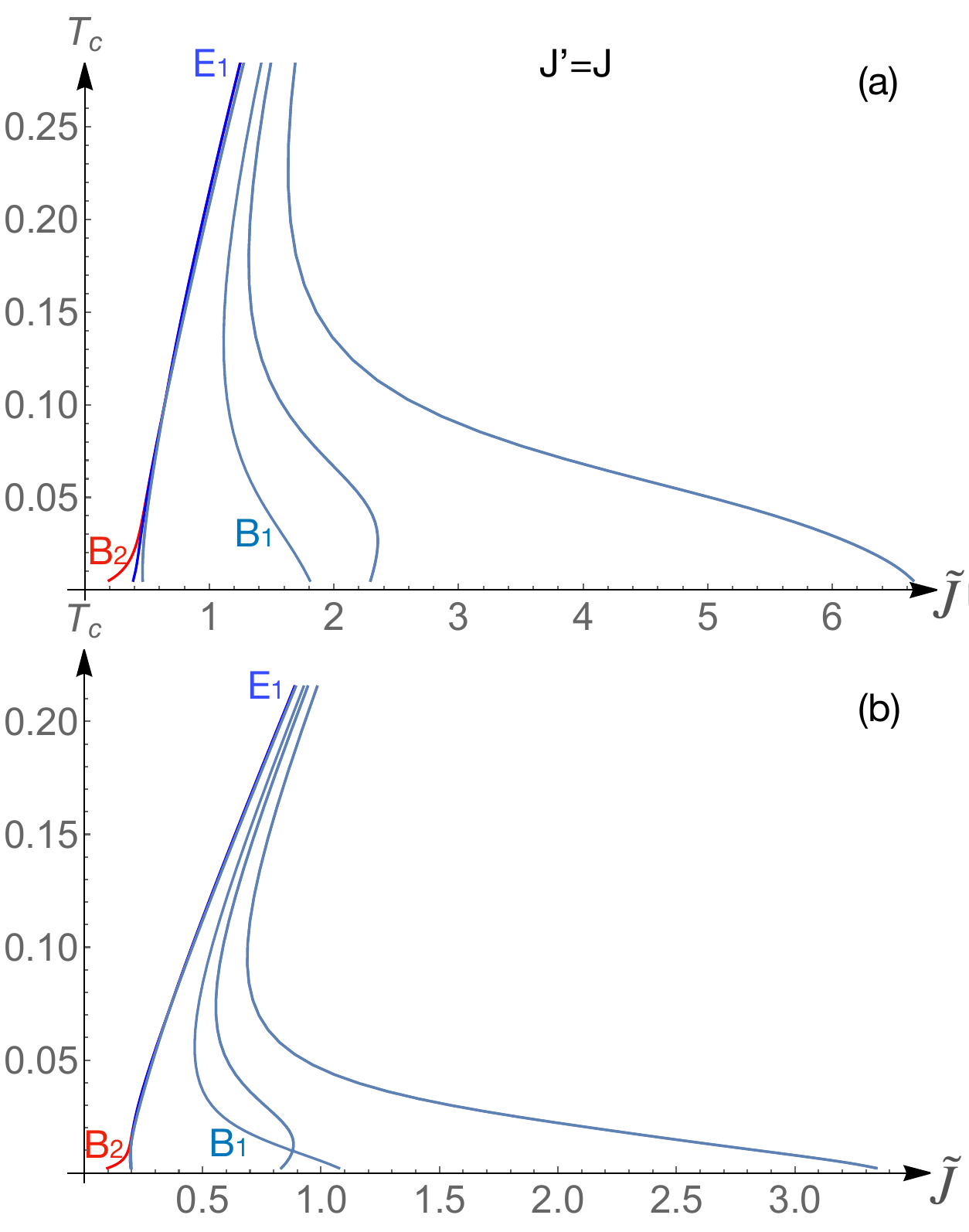}
%	%%includegraphics[width=5cm]{TcpairJpJdop003.pdf}
%	%%includegraphics[width=7.5cm]{ TcpairJpJ(2).pdf}
	\caption{Dependence of the superconducting critical temperatures on the strength of the magnetic coupling, ${\tilde J}$ obtained from the 
	linearized gap equations at $J'=J$ for doping (a) $\delta=0.1$ and (b) $\delta=0.04$. The most favorable solution at a given ${\tilde J}$ occurs for the channel with the highest $T_c$. While the dominant superconducting solution at small coupling,  ${\tilde J} \rightarrow 0$, 
	is $f$-wave (B$_2$; red solid line), at large  ${\tilde J}$, the doubly-degenerate $p$ wave (E$_1$; magenta solid line) solution is found. 	$t = 1$ in these plots. 
	}
	\label{fig:chanJpJ}
\end{figure} 
We can understand the differences between real and reciprocal space pairing by comparing the intra- and intra-band order parameters.
The intraband gap functions, $\Delta_{mm}({\bf k})$, for $f$-wave (B$_2$) solutions for $J'=J$ and $\delta=0.09$ with  $m=2, 3$,  shown in Fig. \ref{fig:gap},
provide a typical example. %Since the $\Delta_{mm}({\bf k})$ are imaginary from Eq. (\ref{eq:gap}) they coincide with their modulus 
The  solution in ${\bf k}$-space projected onto the flat band at the Fermi energy, $m=3$, shows a $p_y$-wave structure (i.e., it is odd under $k_y\rightarrow -k_y$, but even under $k_y\rightarrow -k_x$) similar to the lowest energy BdG dispersion, Fig. \ref{fig:BdGbands}(d). 
However, we also find large that the interband contributions, $|\Delta_{m3}({\bf k})| \lesssim 0.22$ for $m \neq 3$, can be quite large  compared to the intraband contribution $|\Delta_{33}({\bf k})| \sim 0.64$. This is important for understand why the real space and reciprocal space pictures are so different. Even more remarkably, for $p+f$ (E$_1$+B$_1$) superconductivity, $\Delta_{33}({\bf k})=0$,  which means that, the  interband contributions $|\Delta_{m3}({\bf k})|$ are solely responsible  the gap 
in the BdG spectrum, Fig. \ref{fig:BdGbands}(b).

The set of linearized gap  equations at T$_c$ obtained by introducing 
the approximate BdG dispersions, $\omega_m({\bf k})$ in Eq. (\ref{eq:sce}), read
\begin{equation}
\Gamma {\bf {\tilde \Delta} } = {1 \over {\tilde J} } {\bf {\tilde \Delta} },
\label{eq:eigen}
\end{equation}
where ${\bf {\tilde \Delta} } = (  {\tilde J} \Delta_A, {\tilde J}' \Delta_X, {\tilde J} \Delta_B )$. Solving for the eigenvalues of \eqref{eq:eigen}, $1/{\tilde J}^\lambda$, 
we can conveniently obtain the dependence of T$_c$ on ${\tilde J}$ in the different pairing channels: $\lambda$,  
described by the eigenvectors ${\tilde \Delta}_\lambda$ (see Appendix A for details). These eigenvectors can be classified according to the character table \ref{table1} as it should since the matrix $\Gamma$ respects the C$_{6v}$ point group symmetry of the lattice.  

%where, again, ${\bf \Delta}$ is the nine component vector defined by the real space pairing amplitudes of Fig. \ref{fig:pairs}. 
%For triplet pairing $\Delta_{ji}=- \Delta_{ij}$, so the pairing amplitudes are multiplied by a minus sign when going against the direction of the arrows. 
The elements of the $9 \times 9$ matrix $\Gamma$ are 
%\begin{widetext}
\begin{eqnarray}
\Gamma_{pq}&=&{1 \over 2 N_s}{\tilde J_p \over \tilde J } \sum_{m,n,{\bf k}} \tanh({\epsilon_m ({\bf k}) \over 2 k_B T} ) {1 \over \epsilon_m({\bf k}) +\epsilon_n({\bf k})} \notag\\
&&\times \left( \Omega^{*p}_{m n} ({\bf k})
\Omega^{q}_{mn}({\bf k})+\Omega^{p}_{mn}({\bf k})\Omega^{*q}_{mn}({\bf k}) \right).
\label{eq:Gamma}
\end{eqnarray}
%\end{widetext}

\begin{figure}
	\centering
	 \includegraphics[width=8cm,clip=]{ 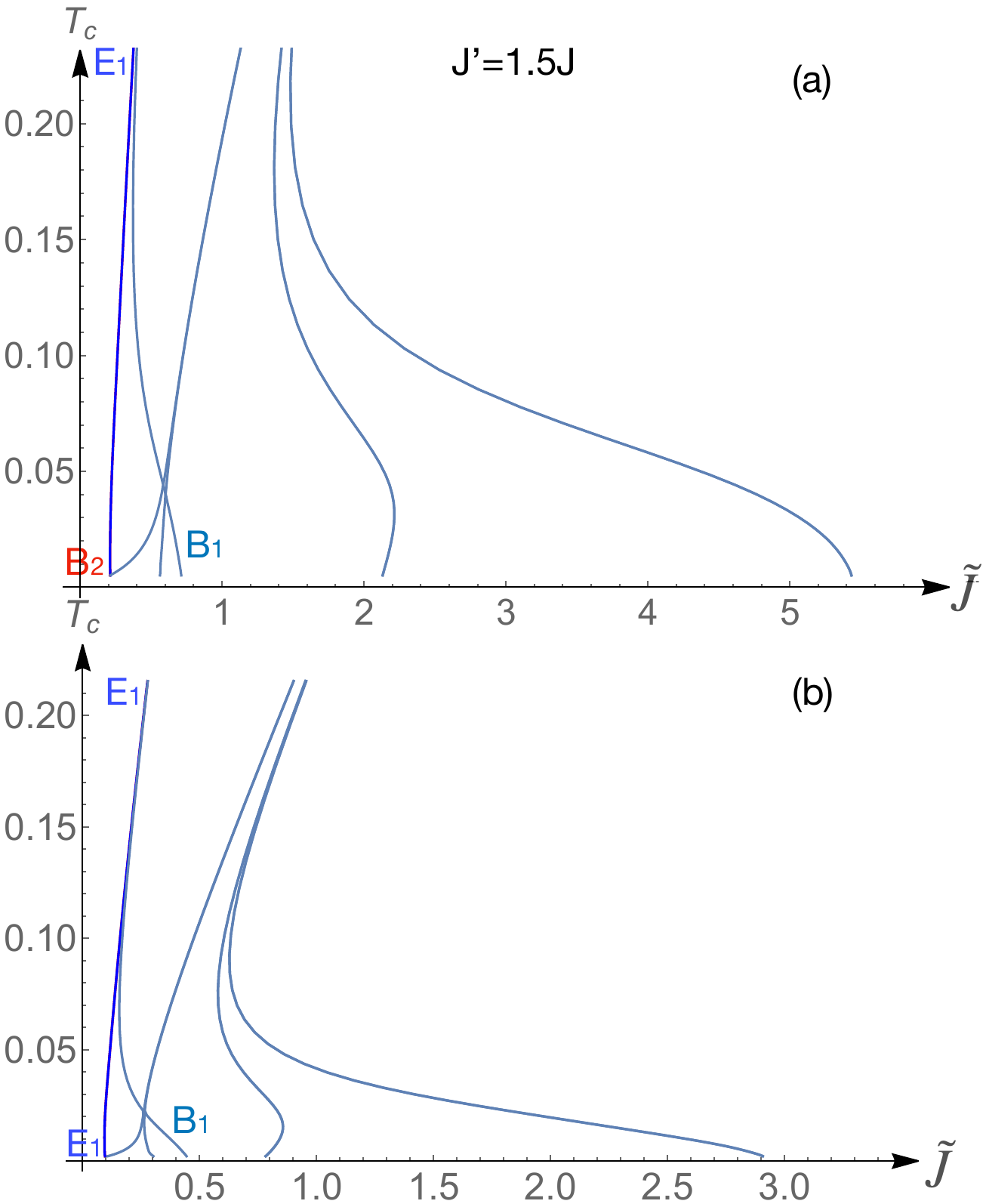}
%	%%includegraphics[width=5cm]{TcpairJpJdop0.pdf}
%	%%includegraphics[width=5cm]{TcpairJpJdop003.pdf}
%	%%includegraphics[width=7.5cm]{ TcpairJp1.5(2).pdf}
	\caption{
	Dependence of the superconducting critical temperatures on  ${\tilde J}$ obtained from the linearized gap equations at $J'=1.5 J$ for doping (a) $\delta=0.1$ and (b) $\delta=0.04$. While the dominant superconducting solution at weak coupling,  ${\tilde J} \rightarrow 0$, the $p$-wave (E$_1$) and $f$-wave (B$_2$) solutions are degenerate within the accuracy of our calculations, for strong coupling the doubly-degenerate $p$-wave (E$_1$; magenta solid line) state has the highest $T_c$. 	$t = 1$ in these plots. 
	}
	\label{fig:chanJp1.5J}
\end{figure}

We can gain insight into the phase diagrams of Fig. \ref{fig:phased} 
obtained from the full SCE by analyzing the dependence of $T_c$ on the strength of the coupling, ${\tilde J}$, as obtained from the linearized gap equation (\ref{eq:eigen}).  The $T_c$s obtained from (\ref{eq:eigen}) for $J'=J$ and $J'=1.5 J$ are shown 
in Figs. \ref{fig:chanJpJ} and  \ref{fig:chanJp1.5J} respectively. From these plots
one can obtain useful information about the dominant pairing channel which is given by the 
one with the largest $T_c$ for a given ${\tilde J}$. 

In the full phase diagrams of Fig. \ref{fig:phased} as $\delta \rightarrow 0$ the effective coupling increases, ${\tilde J}/{\tilde t} 
\propto g_J/g_t \propto 1/\delta$ due to the effect of the GA projection. We 
consistently find that the $f$-wave (B$_2$) solution arises for large doping $\delta \sim 0.1-0.2 $ in the solution of the SCEs indeed 
has the highest $T_c$ in the linearized equations as ${\tilde J} \rightarrow 0$, 
Figs. \ref{fig:chanJpJ}(a) and \ref{fig:chanJp1.5J}(a). As 
${\tilde J}$ increases the $E_1$ has the highest $T_c$. However, the 
$T_c$'s of other solutions become very similar to the $T_c$ of the E$_1$ solution.
This means that several superconducting solutions are numerically indistinguishable. The E$_1$, A$_2$, and B$_2$ solutions have nearly
identical $T_c$s and the $T_c$ of the B$_1$ solution is becomes  close to these 
as ${\tilde J}$ increases. This is consistent with the full non-linear 
SCE which while for $J'=1.5 J$ give a pure E$_1$ solution, 
for $J'=J$ a E$_1$+B$_1$ combination eventually wins.
The SCE also show that for $J'=1.5J$ a complex combination of two E$_1$ 
 is energetically favorable. 
 %This is consistent with results 
%on the doped $t-J$ model on the honeycomb lattice which 
%is known to host\cite{Doniach2007} E$_1$ superconductivity with broken TRS.

\begin{figure}
   \centering
   \includegraphics[width=8cm,clip=0]{ 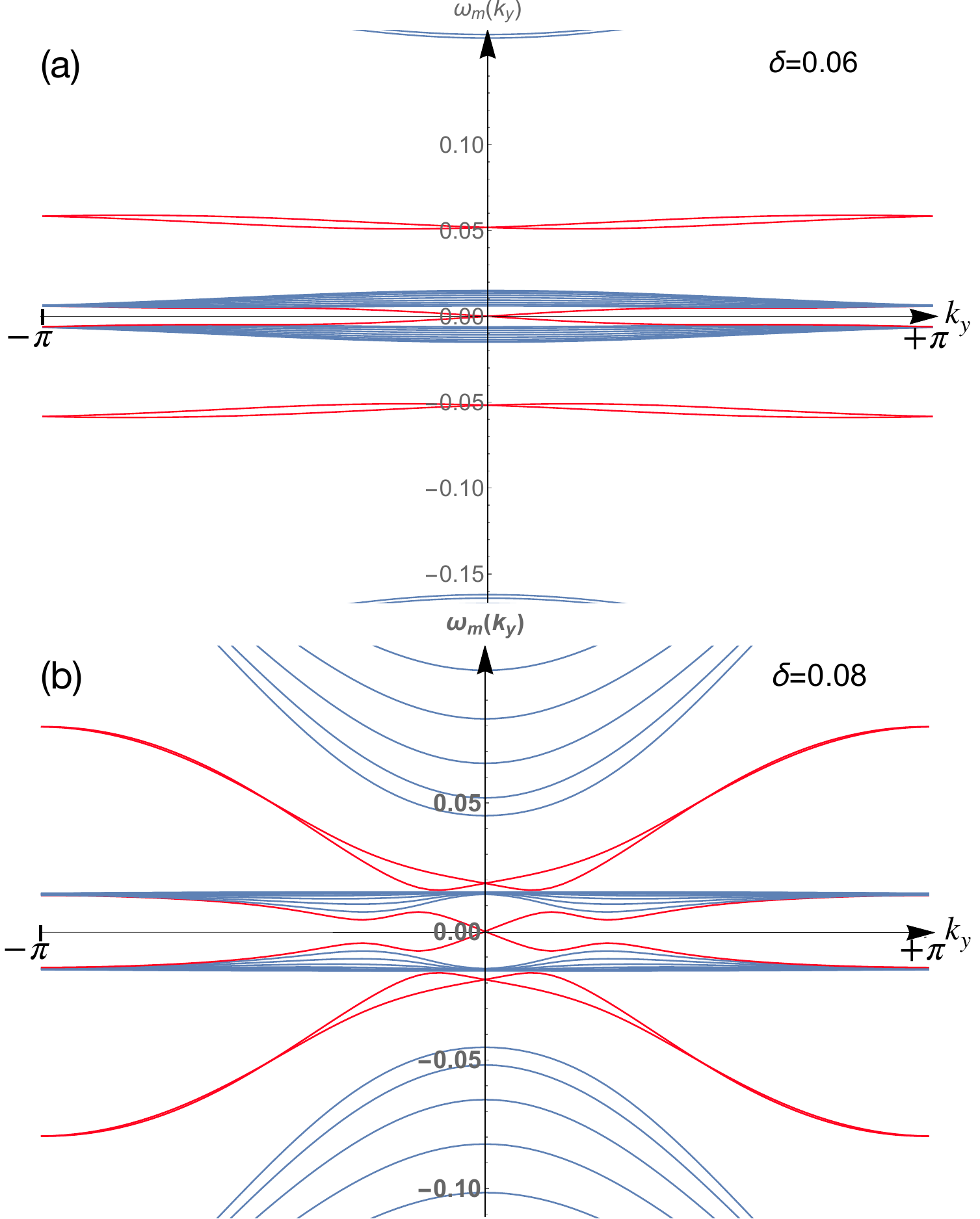}
\caption{Topological edge states in the TSC (E$_1$) phase. The BdG band 
structure of a superconducting ribbon with zig-zag edges along the $y$-direction 
is shown for $J'=1.5J$ and $\delta=0.06$ (top) and $\delta=0.08$ (bottom).
The gapped BdG bands (blue) are traversed by surface edge states
(red) crossing at $k_y=0$.
Edge states are identified by direct comparison with the  spectrum of the extended system. Note that as $t=1$ in these plots, the Majorana modes are extemely localized (i.e., the Majorana bands are extremely flat).
}
 \label{fig:edge}
 \end{figure} 
More details on the analysis of the $T_c$s are provided in Appendix \ref{app::SCTc}. 

The ED pairing correlations 
at $\delta=0$ (see Appendix \ref{app::pairingED})
show that while the E$_1$ and B$_1$ solutions are dominant for $J'/J >1/\sqrt{2}$, the A$_2$ and B$_2$ are favorable for $J'/J< 1/\sqrt{2}$. 
%Differences between the ED and the weak coupling analysis can be attributed to the small six-site clusters analyzed.
These limits are in good agreement with the present weak coupling analysis, considering the small six-site cluster analysed with ED. 

%For $J'=1.5J$ and small $g/t$ the largest $T_c$ corresponds to the B$_2$ and E$_1$ solutions as shown in Fig. \ref{fig:chanJp1.5J}. On the other hand, 
%at large $g/t$ coupling the $T_c$ of the E$_1$ solution becomes largest.  
%The linearized equations do not distinguish among all possible linear combinations of E$_1$ solutions. Indeed, the SCEs selects the complex combination of the two degenerate E$_1$ superconducting solutions shown in the phase diagram of Fig. \ref{fig:phased}.
%Although the B$_1$ solution \textcolor{red}{what does this mean?} the E$_1$ at large couplings, the B$_1$ is not part of the final superconducting combination of solutions as found for $J'=1.5J$.
%Insight into this solution can be gained from exploring the large $J'/J$ limit of our model on the DHL. In the $J'/J \gg 1$ and $\delta \rightarrow 0$ limit, the magnetic exchange of the DHL becomes that of the honeycomb lattice. The doped $t-J$ model on the honeycomb lattice is known to host\cite{Doniach2007} E$_1$ superconductivity with broken TRS consistent with our results. 
%The competition  
%between the $TSC(E_1)$ and the $SC(E_1+B_1)$ states 
%at large $J'/J$ suggested by Fig. \ref{fig:TSCJ} is  evident in 
%the weak coupling analysis shown in
%Fig.\ref{fig:chanJp1.5J} in which the 
%B$_1$ solution merges with the dominant 
%$E_1$ at a lower $T_c$ as $J'/J$ increases.
 \section{Topological superconductivity}
 \label{sec::TSC}
Unambiguous signatures of non-trivial topology are found in both the presence  
of topological edge states and a non-zero topological bulk invariant,
as guaranteed by the bulk-boundary correspondence. In the case of time reversal symmetry breaking superconductors the topological invariant is the Chern number $\nu$ which we  evaluated as described in Appendix \ref{app::Chern}. 
We find that the $p+ip$ (E$_1$) TSC phase has a Chern number, $\nu=1$,
while the other phases discussed are non-topological, $\nu=0$.

From the Chern number calculation we conclude that the $p+ip$ (E$_1$) TSC has a single Majorana edge mode circulating around the edge of the sample. We obtain the edge states of a ribbon in the zig-zag geometry cut along the $y$-direction assuming the lattice orientation shown in Fig. \ref{fig:cluster}. The ribbon is assumed to be in a superconducting state with 
pairing amplitudes given by the bulk solution,  i.e.,
we do not perform a fully self-consistent calculation of the ribbon.
We plot the projected BdG bands along the $k_y$ edge
of a ribbon in the $p+ip$ (E$_1$) TSC state in Fig. \ref{fig:edge}. As well as the bulk bands, two edge states 
with opposite slopes (velocities) cross at $k_y=0$ inside the gap around zero energy. These correspond to the chiral edge state circulating
in opposite directions on the two edges of the ribbon as expected for $\nu=1$. In contrast, no topological edge states are found in ribbons in the $f$-wave (B$_2$) and $p+f$ (E$_1$+B$_1$) superconducting phases as expected for $\nu=0$. Interestingly, the Majorana edge modes associated with the 
$p+ip$ (E$_1$) TSC are almost localized since it emerges within a small gap in nearly flat bands. 

\begin{figure}
	\centering
	\includegraphics[width=8.2cm]{ 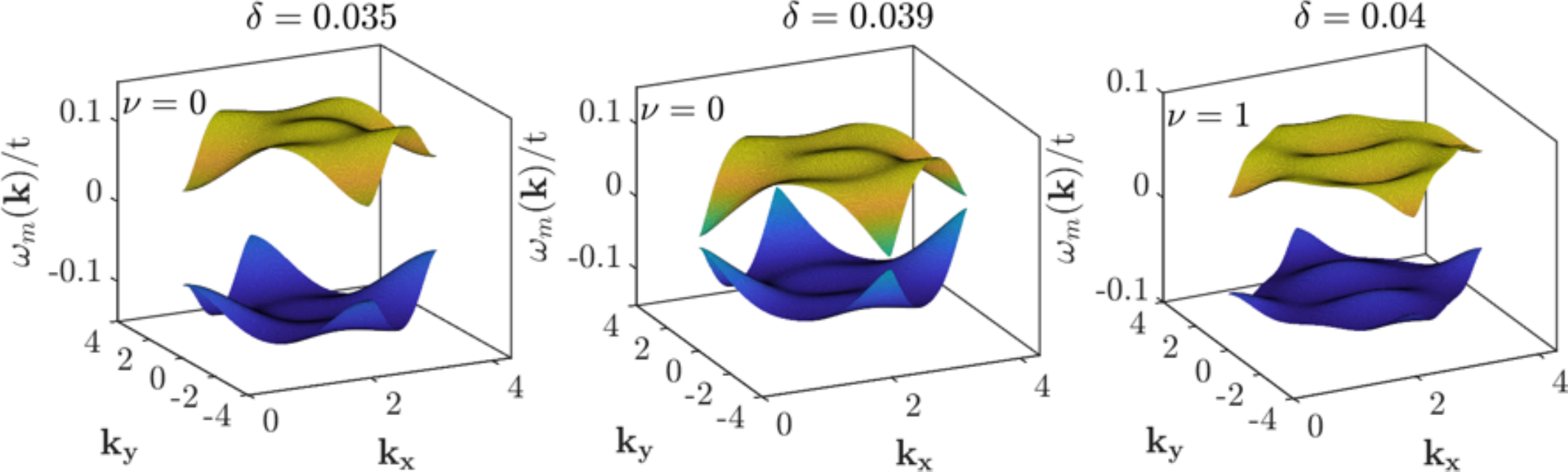}
	\caption{Dependence of the BdG bands, $\pm\omega_{1}({\bf k})$ on doping, $\delta$, across the low doping 
	topological transition for $J'/J=1.5$. At $T=0.01$ as $\delta$ increases from $\delta=0$ a topological gap with $\nu=1$ opens around $\delta_c\sim 0.385$ signalling the TSC(E$_1$) phase of Fig. \ref{fig:phased}.}
	\label{fig:top_trans}
\end{figure}
A topological transition from the trivial $p+ip$ (E$_1$) superconductor to the
topological $p+ip$ (E$_1$) superconductor occurs at $\delta_c \sim 0.039$ 
in the phase diagram \ref{fig:phased} for $J'=1.5J$. The smooth closing and reopening of the superconducting gap (see Fig. \ref{fig:top_trans}) is accompanied by the Chern number jump,  $\Delta \nu=1$, which indicates a second order topological transition at $\delta_c$. At a larger $\delta$ ($\delta \approx 0.14$ at $T=0.01t$), a first order transition from the $p+ip$ (E$_1$) TSC to the non-topological $f$-wave (B$_2$) phase occurs. This contrasts with the doped Kitaev model\cite{Ashwin2012} in which only first order transitions between superconducting states occur. At temperatures above the condensation temperature, $T>T_{\text{BEC}}$, the second order transition at $\delta_c$ becomes a transition from a tRVB to a {\it chiral} tRVB quantum spin liquid (see Fig. \ref{fig:phased}). 

\section{Conclusions}
\label{sec::conclusion}

In this work we have found unconventional triplet superconductivity in an 
easy-axis XXZ ferromagnetic $t$-$J$ model on the DHL. The 
underlying mechanism is attributed to 
the preexistence of a tRVB spin liquid in the Mott phase which is 
suggested by numerical exact diagonalization calculations. 
The symmetry of  the superconducting order parameter depends sensitively 
on both  $J'/J$ and doping. This  is a direct consequence of the flat band, which favors many superconducting channels.

At small $\delta$ 
ferromagnetic interactions dominate leading to exotic phases including 
topological superconductivity.  However, at large $\delta$ superconductivity is mainly
weak coupling, and strongly influenced by band structure effects. A similar situation 
including topological superconductivity was
predicted for hole doped Kitaev spin liquids with ferromagnetic exchange interactions.\cite{Ashwin2012}

Generically, lines of
first order transitions 
separate  different superconducting phases. The important exception to this is the transition from topologically trivial $p+ip$ (E$_1$) superconductivity to topological $p+ip$ (E$_1$) superconductivity with Chern number, $\nu=1$. This transition is continuous, which is possible as both states transform according to the same irreducible representation of $C_{6v}$. This occurs in a two-dimensional representation is, perhaps, unsurprising in retrospect, as  two-dimensional representations lead quite naturally to broken time reversal symmetry \cite{Powellgroup}.

Topological superconductivity occurs in a broad doping range
of the phase diagram being favored at 
large $J'/J$. Critical 
temperatures of $T_c \sim 0.8 J$ are reached at
optimal hole doping, $\delta \sim 0.06$ for $J'=1.5J$. 
The corresponding Majorana edge modes obtained at the 
lowest dopings are almost localized in two flat bands close to the Fermi level. 
This suggests that topological superconductivity in flat band materials could be useful for decoherence free topological quantum computing schemes based on localized Majorana modes.

We have shown that the nodal structures of the superconducting gap appear very different in real space and reciprocal space. This results from the need to transform from a real-space multi-site basis to a reciprocal space multi-band basis. There is nothing special about the DHL lattice in this regard. Therefore, we expect this to be observed in any superconductor where multiple (atomic \cite{Senechal2019} or molecular) orbitals at distinct sites contribute significant density of states near the Fermi energy.

It is also interesting that we observed $p+f$ superconductivity with only a single superconducting transition even though $p$-wave and $f$-wave superconductors belong to different representations of C$_{6v}$ -- the point group describing the DHL, E$_1$ and B$_1$ or B$_2$ respectively. This occurs because the dominant $p$ wave superconducting state lowers the symmetry of the system, such that the $p$-wave and $f$-wave belong to the same irreducible representation, i.e., they become overtones, and arbitrary admixtures are allowed without breaking additional symmetries. That this is possible has long been understood \cite{Sigrist1991,Leggett1975}, however there are still relatively few detailed calculations where this is observed.

A crucial ingredient of our model is the easy-plane 
ferromagnetic interaction which can arise in spin orbit Mott insulators. Although candidate materials such as 
Mo$_3$S$_7$(dmit)$_3$, (EDT-TTF-CONH$_2$)$_6$[Re$_6$Se$_8$(CN)$_6$], and Rb$_3$TT$\cdot$2H$_2$O host strong correlations 
on a DHL antiferromagnetic interactions are dominant. Increasing SMOC in these candidate materials, for example by synthesizing the Se analogue of Mo$_3$S$_7$(dmit)$_3$ or finding other spin orbit Mott insulators on
DHLs are routes for realizing  triplet superconductivity under hole doping. 
Hole doping has been achieved in certain organics\cite{Kanoda2017} but only at a specific doping fixed by the chemistry of the constituents. Clearly, 
experimental progress in hole doping techniques are required to enlarge hole doping ranges in 
organic and organometallic materials and test our prediction of triplet topological superconductivity in the DHL.

\acknowledgments
We acknowledge financial support from (RTI2018-098452-B-I00) MINECO/FEDER, Uni\'on Europea and the Australian Research Council (DP180101483).

\appendix

\section{Superconducting critical temperatures and pairing symmetries}
\label{app::SCTc}
We now analyze the pairing symmetries and $T_c$s based on the weak coupling 
analysis of Section \ref{sec:SCpairing} of the main text.
\begin{figure*}
   \centering
  \includegraphics[width=17cm,clip=]{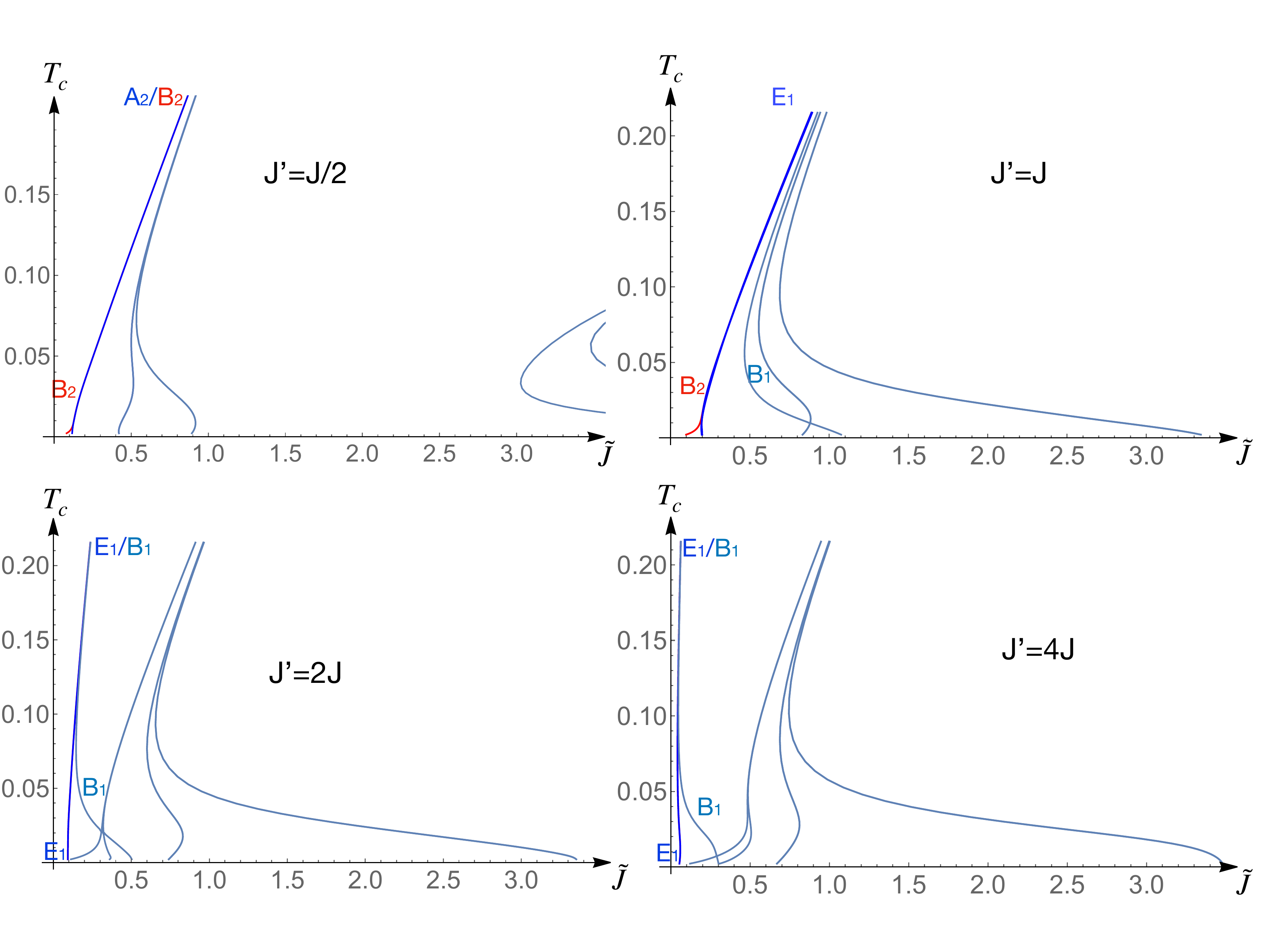}
%%%%includegraphics[width=8cm]{TcpairJp05Jdop0.pdf}
 %%%%includegraphics[width=8cm]{TcpairJp2Jdop0.pdf}
\caption{
 Dependence of the critical temperatures in different pairing channels  on the strength of the magnetic coupling, ${\tilde J}$, for different ratios $J'/J$ at $\delta=0.04$  obtained from the linearized gap equations. The most favorable solution at a given ${\tilde J}$ occurs for the channel with the highest $T_c$.
}
 \label{fig:Tc}
 \end{figure*} 
The matrix elements, $\Omega^{p}_{mn}({\bf k})$ in Eq. (\ref{eq:Gamma})
read:
\begin{eqnarray*} 
\Omega^1_{mn}({\bf k} )&=&C_{A1}^*(\epsilon_m) C_{A2}(\epsilon_n) +C_{A2}^*(\epsilon_m) C_{A1}(\epsilon_n)
\nonumber \\
\Omega^2_{mn}({\bf k} )&=&C_{A1}^*(\epsilon_m) C_{A3}(\epsilon_n) +C_{A3}^*(\epsilon_m) C_{A1}(\epsilon_n)
\nonumber \\
\Omega^3_{mn}({\bf k} )&=&C_{A2}^*(\epsilon_m) C_{A3}(\epsilon_n) +C_{A3}^*(\epsilon_m) C_{A2}(\epsilon_n)
\nonumber \\
\Omega^4_{mn}({\bf k} )&=&C_{A1}^*(\epsilon_m) C_{B1}(\epsilon_n) e^{i {\bf k} {\delta}_1} +C_{B1}^*(\epsilon_m) C_{A1}(\epsilon_n) e^{-i {\bf k} {\delta}_1}
\nonumber \\
\Omega^5_{mn}({\bf k} )&=&C_{A2}^*(\epsilon_m) C_{B2}(\epsilon_n)e^{i {\bf k} {\delta}_2 }+C_{B2}^*(\epsilon_m) C_{A2}(\epsilon_n)e^{-i {\bf k} {\delta}_2}
\nonumber \\
\Omega^6_{mn}({\bf k} )&=&C_{A3}^*(\epsilon_m) C_{B3}(\epsilon_n) e^{i {\bf k} {\delta}_3}+C_{B3}^*(\epsilon_m) C_{A3}(\epsilon_n)e^{-i {\bf k} {\delta}_3}
\nonumber \\
\Omega^7_{mn}({\bf k} )&=&C_{B1}^*(\epsilon_m) C_{B2}(\epsilon_n) +C_{B2}^*(\epsilon_m) C_{B1}(\epsilon_n)
\nonumber \\
\Omega^8_{mn}({\bf k} )&=&C_{B1}^*(\epsilon_m) C_{B3}(\epsilon_n) +C_{B3}^*(\epsilon_m) C_{B1}(\epsilon_n)
\nonumber \\
\Omega^9_{mn}({\bf k} )&=&C_{B2}^*(\epsilon_m) C_{B3}(\epsilon_n) +C_{B3}^*(\epsilon_m) C_{B2}(\epsilon_n),
\label{eq:Omega}
\end{eqnarray*}
where the $C_i(\epsilon_m)$ are the Bloch wave coefficients of the $m$-band:
\begin{equation}
|\Psi_m({\bf k})\rangle = {1 \over \sqrt{N_c} } \sum_{\alpha, i} C_{\alpha i}(\epsilon_m) \sum_n e^{i {\bf k} \cdot ( {\bf R}_n+{\bf d}_{\alpha i}) }  | \phi_{n \alpha i} \rangle,   
\end{equation}
where $| \phi_{n \alpha i} \rangle = | \phi({\bf r} - {\bf R}_n - d_{\alpha i} ) \rangle$. The symmetries of the pairing amplitudes are determined by these coefficients which carry the information of the lattice symmetries. 
From the eigenvalues of $\Gamma$, ${ 1 \over {\tilde J}_\lambda  }$, we obtain the Cooper pair formation energy, ${\tilde J}_\lambda$, in each channel pairing  $\lambda$.
The smaller the ${\tilde J}_\lambda$, the easier it is
to form Cooper pairs in the corresponding channel. Hence, we can determine the dominant pairing channels allowed by symmetry given in Table \ref{table1}.

We illustrate in detail the $J'=J$ case which is amenable to analytical analysis.  
The structure of the $\Gamma$ matrix in Eq. (\ref{eq:Gamma}) is:
\begin{equation*}
\Gamma=
\begin{pmatrix}
A_{3 \times 3} & B_{3 \times 3}  & C_{3 \times 3}  \\
B^\intercal_{3 \times 3} & D_{3 \times 3} & -B^\intercal_{3 \times 3} \\
C_{3 \times 3} & -B_{3 \times 3}  & A_{3 \times 3}.  \\
\end{pmatrix}
 \end{equation*}
where:
\begin{eqnarray}
A_{3 \times 3} = \left( \begin{array}{ccc}
a_1 & a_2 & a_2 \\
a_2 &  a_1 & a_2 \\
a_2 & a_2 & a_1 \\
\end{array} \right),\\
B_{3 \times 3} = \left( \begin{array}{ccc}
0 & b_1 & -b_1 \\
-b_1 & 0 & b_1 \\
b_1 & -b_1 & 0 \\
\end{array} \right),
\\
C_{3 \times 3} = \left( \begin{array}{ccc}
c_1 & c_2 & c_2 \\
c_2 &  c_1 & c_2 \\
c_2 & c_2 & c_1 \\
\end{array} \right),
\end{eqnarray}
and
\begin{eqnarray}
D_{3 \times 3} = \left( \begin{array}{ccc}
d_1 & d_2 & d_2 \\
d_2 & d_1 & d_2 \\
d_2 & d_2 & d_1 \\
\end{array} \right).
\end{eqnarray}
%For simplicity we have defined: $a_1=\Gamma_{11}, a_2=\Gamma_{12}, b_1=\Gamma_{15}, c_2=\Gamma_{18}, d_2=\Gamma_{45} $.
From the common eigenvectors: $\phi_{sym}$, of the $A, B, C, D$ matrices 
and their corresponding eigenvalues: $\tilde{a}_{sym}, \tilde{b}_{sym}, \tilde{c}_{sym}$ and $\tilde{d}_{sym}$ we can find solutions analytically. Introducing: $ [ \Delta_\lambda ]^\intercal = (\phi_{sym},(0,0,0),\pm \phi_{sym})$ in Eq. (\ref{eq:eigen}) leads to the A$_2$ and B$_2$ 
solutions:
\begin{align}
[ \Delta(\text{A}_2) ]^\intercal & \propto[ (1,1,1),(0,0,0), (1,1,1) ]
\nonumber \\
[ \Delta(\text{B}_2) ]^\intercal & \propto [(1,1,1),(0,0,0), (-1,-1,-1) ]
\end{align}
with corresponding eigenvalues: 
\begin{align}
{1 \over {\tilde J}_{\text{A}_2}} &= \tilde{a}_{sym} + \tilde{c}_{sym}=a_1+2a_2 + (c_1+2c_2), 
\nonumber \\
{ 1 \over {\tilde J}_{\text{B}_2}} &= \tilde{a}_{sym} - \tilde{c}_{sym}=a_1+2a_2 - (c_1+2c_2). 
\end{align}
There are also two degenerate solutions of the E$_2$ type: 
\begin{align}
[ \Delta(\text{E}_2) ]^\intercal & \propto[ (-1,0,1),(0,0,0), (-1,0,1) ]
\nonumber \\
[ \Delta(\text{E}_2) ]^\intercal & \propto [(1,-1,0),(0,0,0), (1,-1,0) ]
\end{align}
with the common eigenvalue: 
\begin{equation}
{1 \over {\tilde J}_{\text{E}_2} } = a_1-a_2 + c_1-c_2.
\end{equation}
Finally, we find the B$_1$ solution:
\begin{equation}
[ \Delta(\text{B}_1) ]^\intercal \propto [(0,0,0),(1,1,1), (0,0,0)] ,
\end{equation}
with eigenvalue
\begin{equation}
{1 \over {\tilde J}_{\text{B}_1}}=d_1+2 d_2. 
\end{equation}
So we have found five solutions of the type 
$ [ \Delta_\lambda ]^\intercal = (\phi_{sym},(0,0,0),\pm \phi_{sym})$.
There are four more solutions of the E$_1$ type:
\begin{align*}
[ \Delta(\text{E}_1^-) ]^\intercal & \propto ((1,0,-1),(\beta,-2\beta,\beta), (-1,0,1) )
\nonumber \\
[ \Delta(\text{E}_1^-) ]^\intercal & \propto ((1,-1,0),(-\beta,-\beta,2\beta), (-1,1,0) )
\nonumber \\
[ \Delta(\text{E}_1^+) ]^\intercal & \propto ((1,0,-1),(-\alpha,2\alpha,-\alpha), (-1,0,1) )
\nonumber \\
[ \Delta(\text{E}_1^+) ]^\intercal & \propto ((1,-1,0),(\alpha,\alpha,-2\alpha), (1,-1,0) )
\end{align*}
where $\alpha \ne \beta$ in general. The above four states have corresponding eigenvalues:
\begin{widetext}
\begin{align*}
{ 1 \over {\tilde J}_{\text{E}_1^-}} &= {1 \over 2} \left(a_1 - a_2 - c_1 + c_2 + d_1 - d_2 -  \sqrt{ (a_1 - a_2 - c_1 + c_2 + d_1 - d_2)^2  + 24 b_1^2 -4 (a_1-a_2-c_1+c_2)(d_1-d_2)} \right)
\nonumber \\
%{1 \over J} &= 1/2 \left(a_1 - a_2 - c_1 - c_2 + d_1 - d_2 -  \sqrt{ (-a_1 + a_2 + c_1 + c_2 - d_1 + d_2)^2  + 24 b_1^2 -4 (a_1-a_2-c_1-c_2)(d_1-d_2)} \right)
%\nonumber \\
{1 \over {\tilde J}_{\text{E}_1^+}} &= {1 \over 2} \left(a_1 - a_2 - c_1 + c_2 + d_1 - d_2 +  \sqrt{ (a_1 - a_2 - c_1 + c_2 + d_1 - d_2)^2 + 24  b_1^2 -4 (a_1-a_2-c_1+c_2)(d_1-d_2)} \right)
\end{align*}
\end{widetext}

In Fig. \ref{fig:Tc} we show the dependence of $T_c$ on the coupling strength
for different anisotropies, $J'/J=0.5,1,2$, and 4. We first discuss the results 
at weak coupling, $ {\tilde J} \rightarrow 0$, which would correspond to large 
$\delta$ regime of Fig. \ref{fig:phased} due to the GA projection. The 
$T_c$ of the B$_2$ solution (red solid line) is the highest for 
small ${\tilde J}$ when  $J',J=0.5, 1$ or 1.5 (the weak coupling 
results for $J'=1.5 J$ and shown in Fig. \ref{fig:chanJp1.5J}).
However, for $J'=2J$, the E$_1$ solution becomes more stable than the 
B$_2$ as ${\tilde J} \rightarrow 0$. These results indicate that the
B$_2$ solution is stable in a broad range of $J'/J$ consistent with the 
full phase diagrams of Fig. \ref{fig:phased}.

On the other hand, as ${\tilde J}$ increases, other solutions become favoured. 
For $J'=J/2$ the $T_c$s of the B$_2$ and the A$_2$ merge so that 
both solutions are favorable, for $J'=J$ the E$_1$ is singled out 
having the highest $T_c$ among all solutions. We also note how the $T_c$
of the B$_1$ becomes the next most highest. In fact, for $J'/J=2, 4$, the $T_c$ 
of the B$_1$ converges to the $T_c$ of the E$_1$ within the ranges of the plot. 
This analysis is 
consistent with the E$_1$ and E$_1$+B$_1$ solutions found at small doping, $\delta$ 
in the phase diagrams of Fig. \ref{fig:phased} and the competition between these 
solutions as $J'/J$ is varied shown in the phase diagrams of Fig. \ref{fig:TSCJ}. 

%The ED calculations of pairing correlations on small clusters of Appendix \ref{app::pairingED} show that while for $ J' \sim 1/\sqrt{2} J$ the A$_2$, B$_2$ and E$_2$ solutions become dominant as $J'/J$ is increased. These results are consistent with the suppression of the weak 
%coupling SC(B$_2$) phase with $J'/J$ observed in the phase diagrams of 
%Fig. \ref{fig:phased}. 

\section{Exact diagonalization analysis}
\label{app::EDanalysis}
Here we give some details on the exact diagonalization calculations of 
model \eqref{eq:modeltJ} at half-filling on small clusters. 
The ground state wavefunction and energy is compared with the 
$|\text{nn-tRVB} \rangle $  [Eq. (\ref{eq:tRVB})]. On the six-site cluster we have 20 possible spin configurations:
\begin{eqnarray}
|1\rangle&=&c^\dagger_{1\uparrow}c^\dagger_{2\uparrow}c^\dagger_{3\uparrow}c^\dagger_{4\downarrow}c^\dagger_{5\downarrow}c^\dagger_{6\downarrow}|0\rangle
\nonumber \\
|2\rangle&=&c^\dagger_{1\uparrow}c^\dagger_{2\uparrow}c^\dagger_{3\downarrow}c^\dagger_{4\uparrow}c^\dagger_{5\downarrow}c^\dagger_{6\downarrow}|0\rangle
\nonumber \\
|3\rangle&=&c^\dagger_{1\uparrow}c^\dagger_{2\uparrow}c^\dagger_{3\downarrow}c^\dagger_{4\downarrow}c^\dagger_{5\uparrow}c^\dagger_{6\downarrow}|0\rangle
\nonumber \\
|4\rangle&=&c^\dagger_{1\uparrow}c^\dagger_{2\uparrow}c^\dagger_{3\downarrow}c^\dagger_{4\downarrow}c^\dagger_{5\downarrow}c^\dagger_{6\uparrow}|0\rangle
\nonumber \\
|5\rangle&=&c^\dagger_{1\uparrow}c^\dagger_{2\downarrow}c^\dagger_{3\uparrow}c^\dagger_{4\uparrow}c^\dagger_{5\downarrow}c^\dagger_{6\downarrow}|0\rangle
\nonumber \\
|6\rangle&=&c^\dagger_{1\uparrow}c^\dagger_{2\downarrow}c^\dagger_{3\uparrow}c^\dagger_{4\downarrow}c^\dagger_{5\uparrow}c^\dagger_{6\downarrow}|0\rangle
\nonumber \\
|7\rangle&=&c^\dagger_{1\uparrow}c^\dagger_{2\downarrow}c^\dagger_{3\uparrow}c^\dagger_{4\downarrow}c^\dagger_{5\downarrow}c^\dagger_{6\uparrow}|0\rangle
\nonumber \\
|8\rangle&=&c^\dagger_{1\uparrow}c^\dagger_{2\downarrow}c^\dagger_{3\downarrow}c^\dagger_{4\uparrow}c^\dagger_{5\uparrow}c^\dagger_{6\downarrow}|0\rangle
\nonumber \\
|9\rangle&=&c^\dagger_{1\uparrow}c^\dagger_{2\downarrow}c^\dagger_{3\downarrow}c^\dagger_{4\uparrow}c^\dagger_{5\downarrow}c^\dagger_{6\uparrow}|0\rangle
\nonumber \\
|10\rangle&=&c^\dagger_{1\uparrow}c^\dagger_{2\downarrow}c^\dagger_{3\downarrow}c^\dagger_{4\downarrow}c^\dagger_{5\uparrow}c^\dagger_{6\uparrow}|0\rangle
\nonumber \\
|11\rangle&=&c^\dagger_{1\downarrow}c^\dagger_{2\uparrow}c^\dagger_{3\uparrow}c^\dagger_{4\uparrow}c^\dagger_{5\downarrow}c^\dagger_{6\downarrow}|0\rangle
\nonumber \\
|12\rangle&=&c^\dagger_{1\downarrow}c^\dagger_{2\uparrow}c^\dagger_{3\uparrow}c^\dagger_{4\downarrow}c^\dagger_{5\uparrow}c^\dagger_{6\downarrow}|0\rangle
\nonumber \\
|13\rangle&=&c^\dagger_{1\downarrow}c^\dagger_{2\uparrow}c^\dagger_{3\uparrow}c^\dagger_{4\downarrow}c^\dagger_{5\downarrow}c^\dagger_{6\uparrow}|0\rangle
\nonumber \\
|14\rangle&=&c^\dagger_{1\downarrow}c^\dagger_{2\uparrow}c^\dagger_{3\downarrow}c^\dagger_{4\uparrow}c^\dagger_{5\uparrow}c^\dagger_{6\downarrow}|0\rangle
\nonumber \\
|15\rangle&=&c^\dagger_{1\downarrow}c^\dagger_{2\uparrow}c^\dagger_{3\downarrow}c^\dagger_{4\uparrow}c^\dagger_{5\downarrow}c^\dagger_{6\uparrow}|0\rangle
\nonumber \\
|16\rangle&=&c^\dagger_{1\downarrow}c^\dagger_{2\uparrow}c^\dagger_{3\downarrow}c^\dagger_{4\downarrow}c^\dagger_{5\uparrow}c^\dagger_{6\uparrow}|0\rangle
\nonumber \\
|17\rangle&=&c^\dagger_{1\downarrow}c^\dagger_{2\downarrow}c^\dagger_{3\uparrow}c^\dagger_{4\uparrow}c^\dagger_{5\uparrow}c^\dagger_{6\downarrow}|0\rangle
\nonumber \\
|18\rangle&=&c^\dagger_{1\downarrow}c^\dagger_{2\downarrow}c^\dagger_{3\uparrow}c^\dagger_{4\uparrow}c^\dagger_{5\downarrow}c^\dagger_{6\uparrow}|0\rangle
\nonumber \\
|19\rangle&=&c^\dagger_{1\downarrow}c^\dagger_{2\downarrow}c^\dagger_{3\uparrow}c^\dagger_{4\downarrow}c^\dagger_{5\uparrow}c^\dagger_{6\uparrow}|0\rangle
\nonumber \\
|20\rangle&=&c^\dagger_{1\downarrow}c^\dagger_{2\downarrow}c^\dagger_{3\downarrow}c^\dagger_{4\uparrow}c^\dagger_{5\uparrow}c^\dagger_{6\uparrow}|0\rangle.
\label{eq:config}
\end{eqnarray}
where $\ket{0}$ is the vacuum state. The ground state of the easy-plane ferromagnetic 
model is indeed the expected triplet $S=1, S_z=0$ which
is described as:
\begin{equation}
|\Psi_0 \rangle = \sum_n a_n | n \rangle.
\end{equation}
where the coefficients  are: $A=A_{17}=A_{22}=A_{15}=A_{11}=A_{36}=A_4=0.3165$;
$B=A_{19}=A_{18}=A_{16}=A_{14}=A_{13}=A_{12}=A_9=A_8=A_7=A_5=A_3=A_2= 0.17667$; and $C=A_1=A_{20}=0.1106$. The exact energy of the anisotropic FM model with $J'=J$ is $\epsilon_0(\text{exact})=-3.541287J$. The exact wavefunction compares well 
with the normalized $|\text{nn-tRVB} \rangle$ state whose non-zero 
coefficients are $A=0.3638$ and $B=C=0.1212$  giving the 
overlap $|\langle \text{tRVB} | \Psi_0 \rangle |=0.9747$ quoted in Sec. \ref{sec::exacttriplet}.
   \begin{figure}[t]
   \centering
   \includegraphics[width=8cm]{ 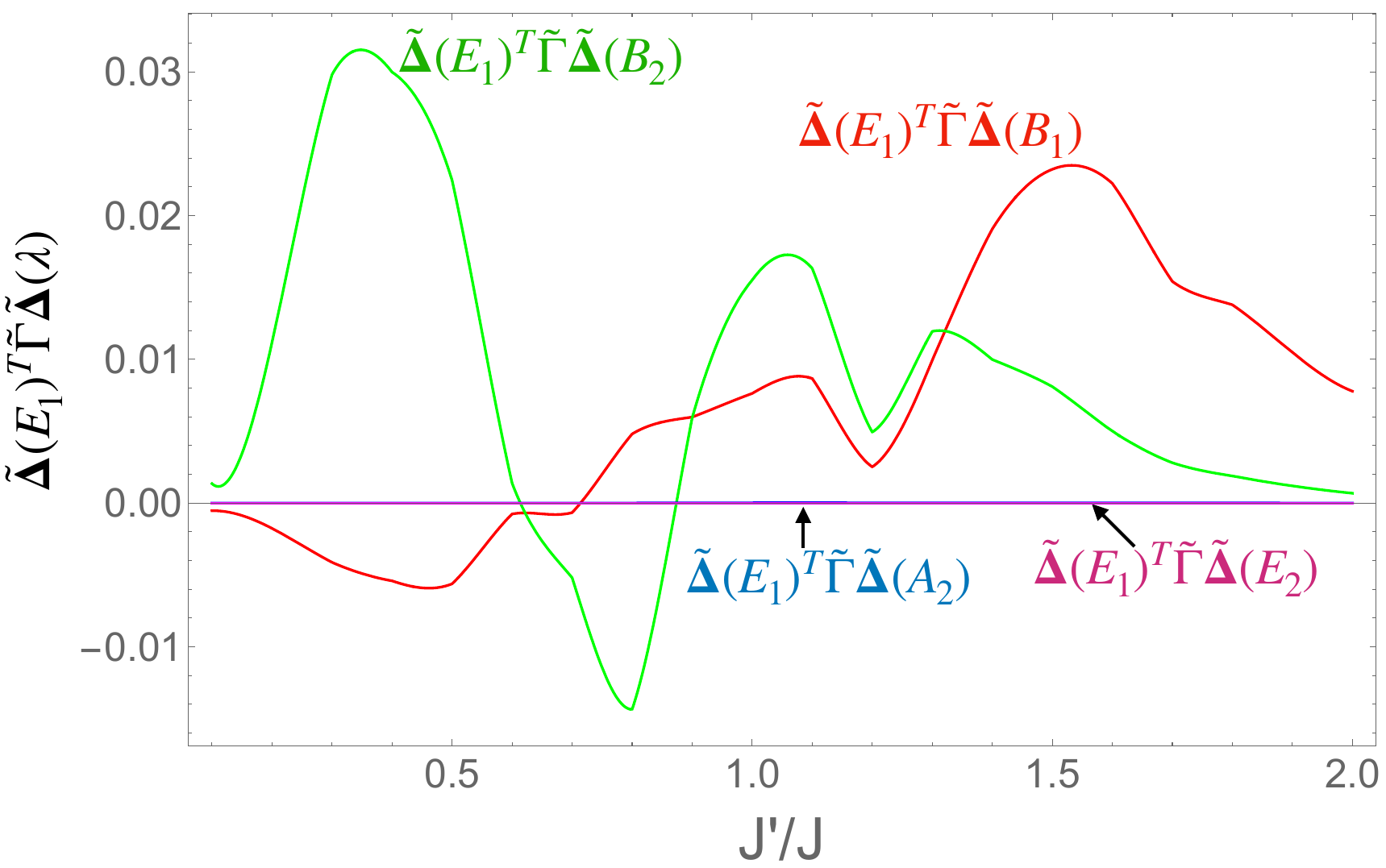}
  \caption{Dependence on $J'/J$ of the hybridization between the gap functions associated with the irreducible representations of C$_{6v}$ on the DHL. The matrix elements of Eq. (\ref{eq:hybrid}) with $\lambda=\text{E}_1$ and $\lambda'=\text{B}_1,\text{B}_2,\text{A}_2$ and E$_2$ are plotted for $\delta=0.04$ and $T/t=0.01<T_c$. Note that only the  E$_1$-B$_1$ and E$_1$-B$_2$ matrix elements are non-zero indicating the possibility of hybrid $p+f$ (E$_1$+B$_1$ and E$_1$+B$_2$) superconducting states, and ruling out $p+i$ (E$_1$+A$_2$) or $p+d$ (E$_1$+E$_2$) superconductivity.}
  \label{fig:hybrid}
  \end{figure}

   \section{Hybrid superconducting states}
\label{app:hybridsc}
As well as the pure pairing states allowed by crystal symmetry, 
the SCE also yield solutions involving more than one 
irreducible representations of the C$_{6v}$ group.
As shown Fig. \ref{fig:phased}, two $p+f$ (E$_1$+B$_1$
and E$_1$+B$_2$) states occur  for $J'=J$. 
Naively one would expect that only solutions
belonging to the same irreducible representation 
of the character table \ref{table1} are allowed. However,
this is only true for the linearized gap equations  (\ref{eq:gap}) at $T_c$ since, in that case,  the
irreducible representations 
are eigenfunctions of the $\Gamma$-matrix. This is because Eq. (\ref{eq:gap}) obeys the full C$_{6v}$ symmetry of the DHL lattice
(since it only depends on the bare band dispersions of the DHL). 

However, for $T<T_c$, i.e., once superconductivity has set in, the C$_{6v}$ symmetry of the system is
spontaneously lowered to a new group, ${\cal G}$, which is a subgroup of C$_{6v}$.
Thus, superconducting orders  associated with different irreducible representations  of C$_{6v}$ may mix, if and only if the are in the same irreducible representation of ${\cal G}$. 
In order to explore possible mixing of superconducting orders we follow \cite{Sigrist1991} and compute the matrix
elements connecting different irreducible representations via  
a modified $\Gamma$ matrix that explicitly contains the symmetry
breaking due to the superconducting order,
%described by a certain $\Delta(\lambda)$ of the
%Table \ref{table1}.
\begin{eqnarray}
{\tilde \Gamma}_{pq}&=&{1 \over 2 N_s} {\tilde J_p \over \tilde J } \sum_{m,n,{\bf k}} \tanh({\omega_m ({\bf k}) \over 2 k_B T} ) {1 \over \epsilon_m({\bf k}) +\epsilon_n({\bf k})} \notag\\
&&\times \left( \Omega^{*p}_{m n} ({\bf k})
\Omega^{q}_{mn}({\bf k})+\Omega^{p}_{mn}({\bf k})\Omega^{*q}_{mn}({\bf k}) \right),
\label{eq:Gammat}
\end{eqnarray}
where $\omega_m({\bf k})$ is given by the 
second order 
expression (\ref{eq:omegam}) with $\Delta_{mn}({\bf k}) $ evaluated with the real space gap function, $\Delta(\gamma)$. Note that this expression 
is appropriate to describe superconducting order  below 
$T_c$ since $\omega_m({\bf k})$ enters this expression in the place of $\epsilon_m$ in
the expression for $\Gamma$ (Eq. (\ref{eq:Gamma})). %\textcolor{red}{Possibly a dumb question - but I just want to check if the $\epsilon_m$'s in the denominator should be $\omega_m$'s too?}

Taking $\Delta(\lambda)$ as the dominant 
gap function, {i.e.}, the one with the highest $T_c$, 
we evaluate matrix elements with the rest of gap functions
of the character table \ref{table1}. These matrix elements read:
\begin{equation}
{\tilde {\bf  \Delta} } (\lambda)^T \tilde {\Gamma} {\tilde {\bf \Delta} }(\lambda').
\label{eq:hybrid}
\end{equation}    
If the matrix element between ${\tilde \Delta} (\lambda)$
and ${\tilde \Delta} (\lambda')$ is non-zero then the 
two pairing solutions are allowed to hybridize below $T_c$. 

  \begin{figure}[t]
   \centering
   \includegraphics[width=8cm]{ 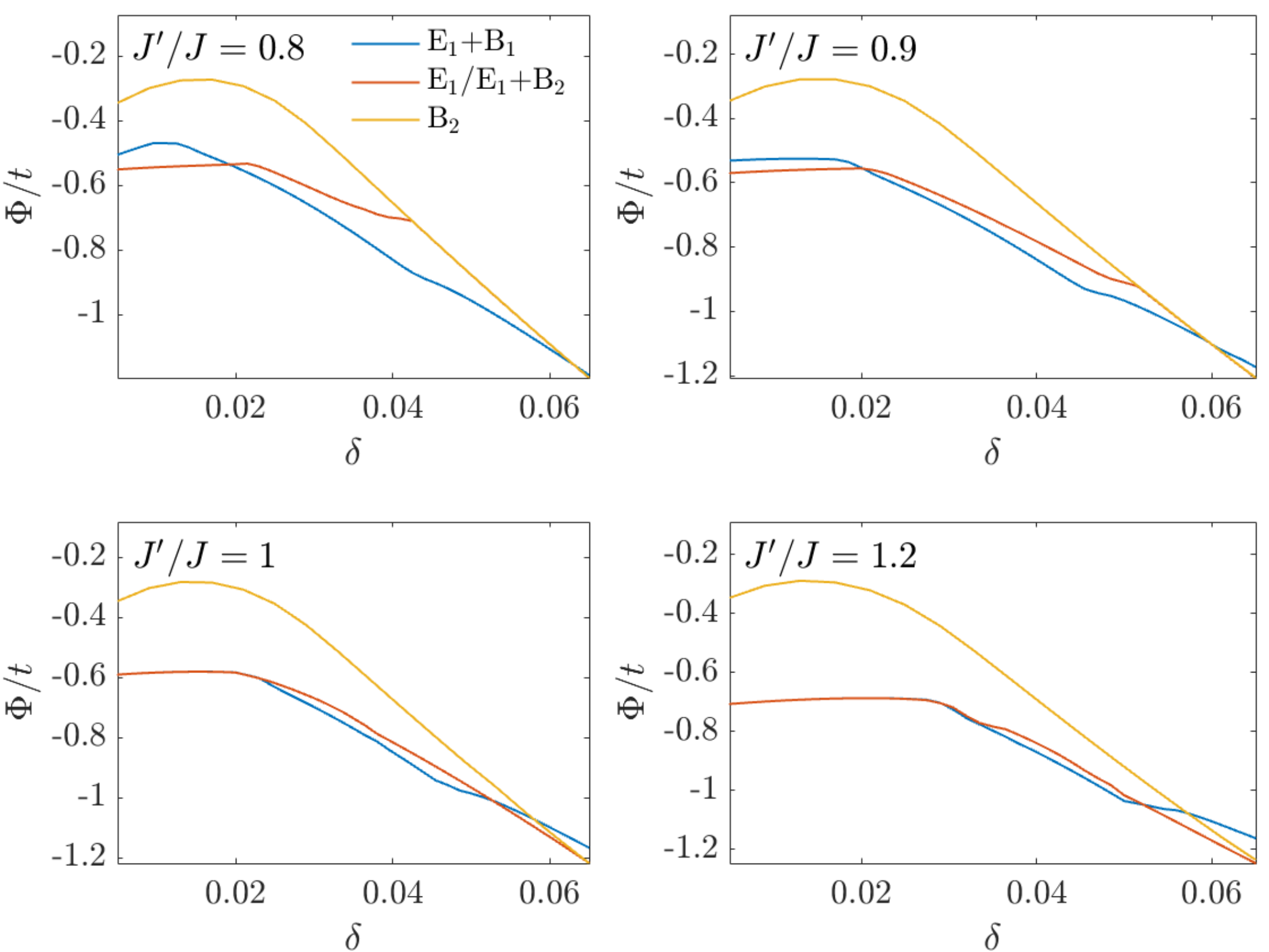}
  \caption{Dependence of the free energies on $\delta$ of the
  most favorable pairing states found in the SCE analysis. }
  \label{fig::Free_energies}
  \end{figure}
 \begin{figure}[t]
   \centering
   \includegraphics[width=8cm]{ 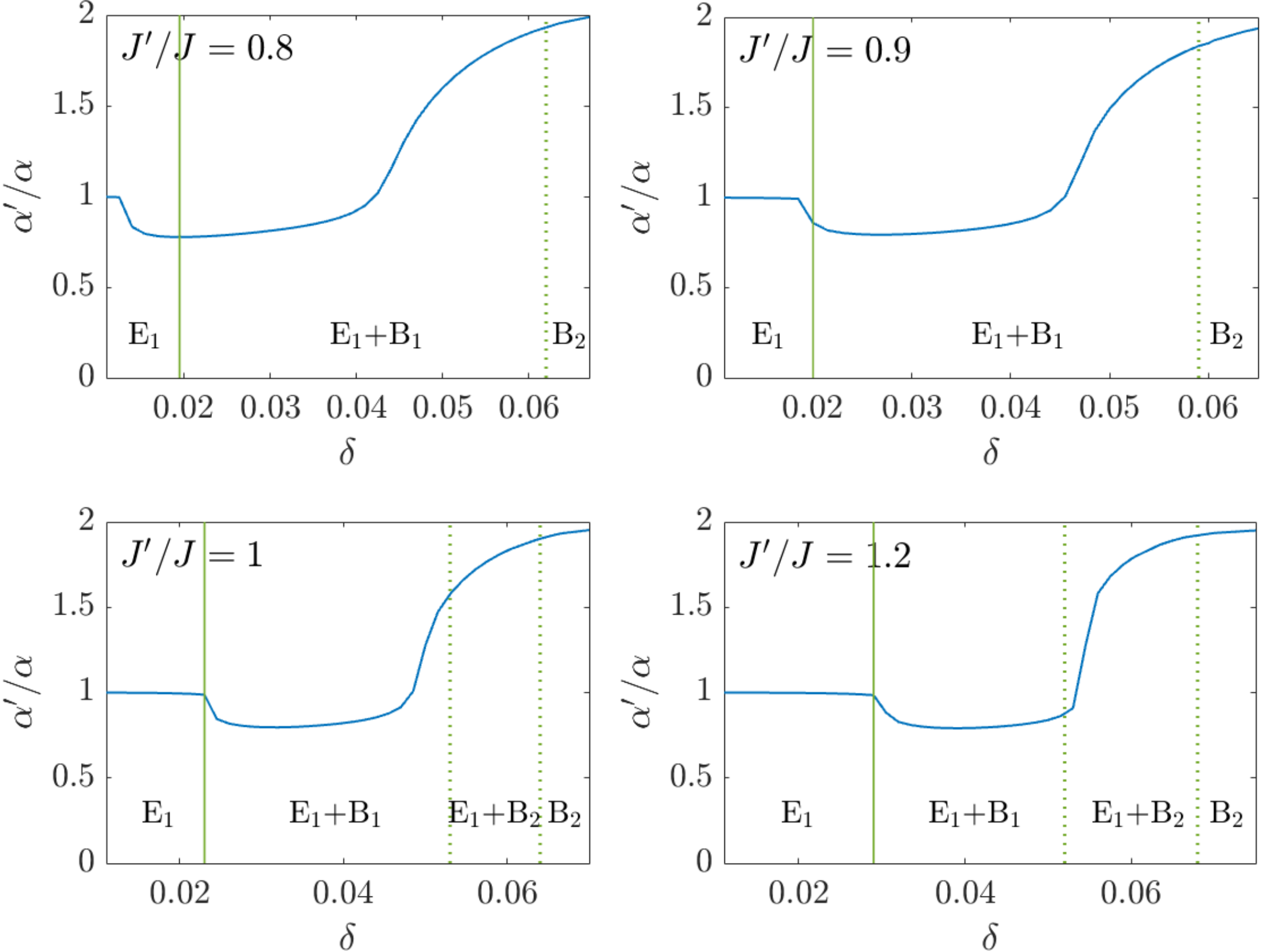}
  \caption{Dependence of the  parameters in the  E$_1$+B$_1$ order parameter (Eq. (\ref{E1B1})), $\alpha'/\alpha$, on $\delta$ for different $J'/J$. The first (dotted green line) and second (solid green line) order transitions separating the  superconducting states are  found from the free energy analysis of Fig. \ref{fig::Free_energies}.}
  \label{fig::aphaE1B1}
  \end{figure}
 \begin{figure}[t]
   \centering
   \includegraphics[width=9cm]{ 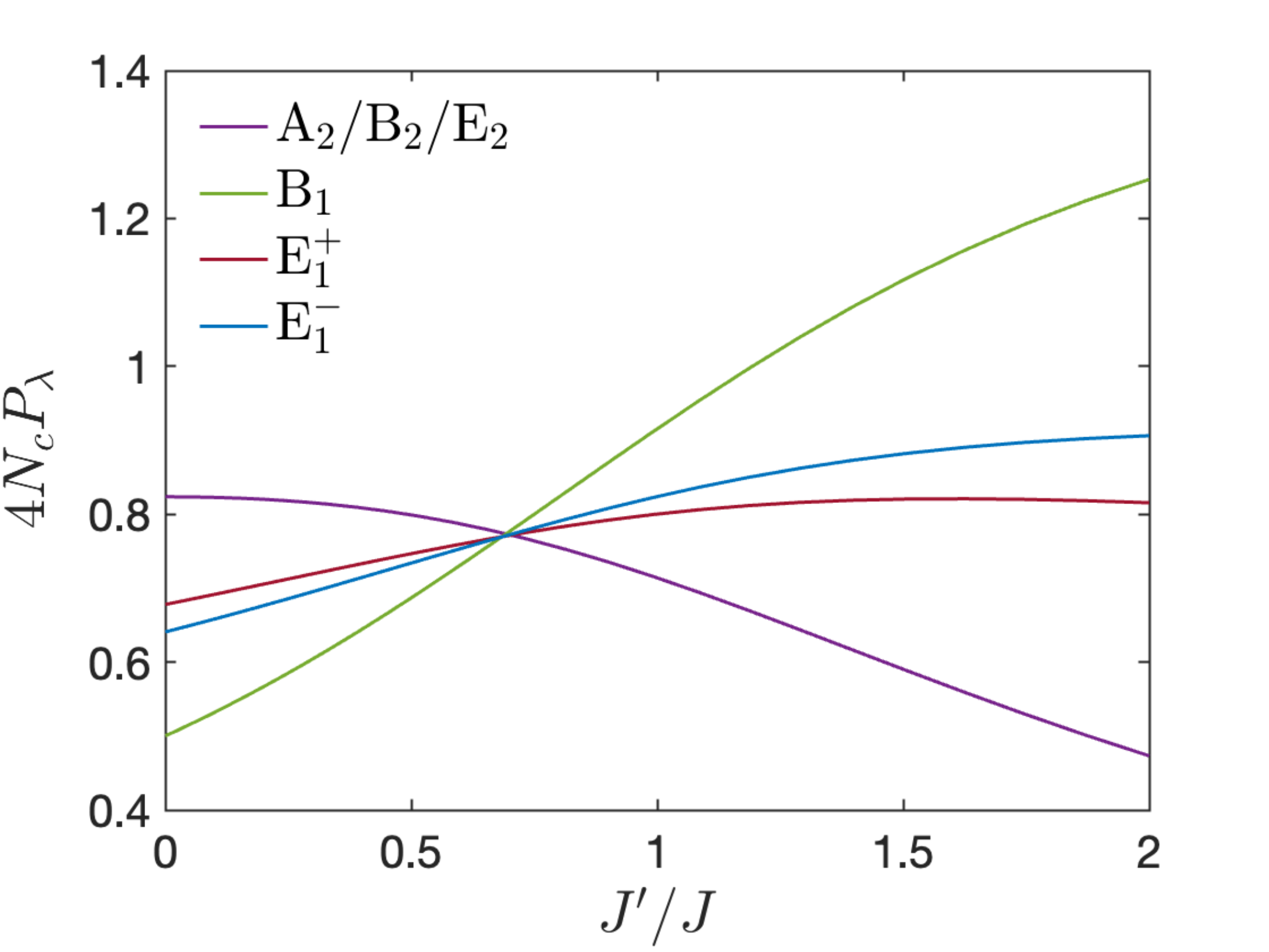}
  \caption{Triplet pairing correlations dependence on $J'/J$ of the half-filled Hubbard model \eqref{eq:Hubbard} on a six-site cluster. Note that the degeneracy of the E$_1^-$ and E$_1^+$ is lifted on the six-site cluster at this has C$_{2v}$ symmetry rather than the C$_{6v}$ symmetry of the full lattice. Nevertheless we retain the C$_{6v}$ labels to retain straightforward connection to the rest of the manuscript.
  %Also remarkable the small size of the lattice considered in these calculations which instead of quantitative differences provides a good qualitative description of the system.
  }
  \label{fig::pairingcorrelations}
  \end{figure} 
As a relevant example to the phase diagrams of Fig. \ref{fig:phased}, we 
evaluate the matrix elements taking $\lambda=\text{E}_1$ since this is the most favorable pure state, i.e., the irreducible representation with the highest $T_c$ for large ${\tilde J}$ for $J'/J=1$ or $1.5$ (see Fig. \ref{fig:chanJpJ} and \ref{fig:chanJp1.5J}). The dependence of matrix elements on $J'/J$ for $T=0.01t$ and a fixed hole doping of $\delta=0.05$ is shown in Fig. \ref{fig:hybrid}. We first note that while
matrix elements between E$_1$ and B$_1$ or B$_2$ states are non-zero, there is no hybridization with A$_2$ and E$_2$ states. This shows that hybrid $p+f$ (E$_1$+B$_1$ or E$_1$+B$_2$) states are allowed below $T_c$ but $p+d$ (E$_1$+E$_2$) and $p+i$ (E$_1$+A$_2$) states are not. This is consistent with the phase diagrams, Fig. \ref{fig:phased}. Since the E$_1$/B$_1$ hybridization is comparable to the E$_1$/B$_2$ around $J'=J$ we can expect both combinations to occur. On the other hand, although around $J'=1.5J$ the E$_1$/B$_1$ hybridization is strong, 
it is the E$_1$ solution that wins in the SCE analysis 
as Fig. \ref{fig:phased} shows. 

The analysis above is then useful to predict which pairing states may combine below $T_c$ due to spontaneous symmetry breaking in the most favorable channel. This can be done before getting
into the solution of the numerically cumbersome SCEs. 

We now provide more details about the hybrid superconducting states obtained from the full SCEs. In Fig. \ref{fig::Free_energies} we compare the 
free energies of the most favorable states around $J'/J=1$ where the 
hybrid superconducting states are stabilised consistent with the weak coupling 
analysis just provided above. For $J'/J=0.8$ and $J'/J=0.9$ the E$_1$+B$_1$ state has the lowest energy in the doping range $0.02 < \delta < 0.06 $ sandwiched between an $E_1$ and a $B_2$. 
On the other hand, for $J'/J=1$ and $J'/J=1.2$, the E$_1$+B$_2$ hybrid state
is most favorable between the E$_1$+B$_1$ and the B$_2$. 
A first order transition from the E$_1$+B$_1$ to the E$_1$ solution occurs at $\delta \sim 0.02$ for $J'/J=0.8, 0.9$, which becomes second order for $J'/J=1,1.2$  consistent 
with the behavior shown in the phase diagrams of Fig.\ref{fig:phased}.

Finally we consider in detail the properties of the E$_1$+B$_1$ state which is the most robust hybrid superconducting state found in the SCE. The dependence of $\alpha'/\alpha$
ratios describing the mixed E$_1$+B$_1$ gap function (Eq. \eqref{E1B1}) is plotted in  Fig. \ref{fig::aphaE1B1} for the same $J'/J$ as shown in  Fig. \ref{fig::Free_energies}.
The smooth variation of $\alpha'/\alpha$ with $\delta$ indicates that the
hybrid E$_1$+B$_1$ state is not an artifact of the numerical solution of the SCE.
Superconducting transitions occur concomitantly with the phase transitions identified from the
free energies plotted in Fig \ref{fig::Free_energies} as expected. 

\section{Pairing correlations from exact diagonalization on small clusters}
\label{app::pairingED}
Following our previous work \cite{Merino2021} we explore unconventional triplet pairing 
using ED techniques on the half-filled Hubbard model with imaginary hopping amplitudes \eqref{eq:Hubbard}. As explained in Sec. \ref{sec::model} this model, at half-filling, maps onto the easy-plane ferromagnetic XXZ  model \eqref{eq:xxz}. We have checked through ED calculations 
that indeed the two models share the same ground state for $U \gg t,t'$.
  
In order to elucidate the pairing symmetry of the tRVB state as 
$\delta \rightarrow 0$  we compute the pairing correlations which can be obtained exactly by evaluating\cite{merino2014}:
\begin{widetext}

\begin{equation}
    P_\lambda=\frac{1}{4N_c}\sum_{\langle \alpha i,\beta j\rangle}\sum_{ \langle \gamma k,\delta l\rangle}\Delta_{\alpha i,\beta j}(\lambda)\Delta_{\gamma k,\delta l}(\lambda)\left(\langle c_{\alpha i\uparrow}^\dagger c_{\beta j \downarrow}^\dagger c_{\gamma k\uparrow}c_{\delta l \downarrow}\rangle -\langle c^\dagger_{\alpha i\uparrow}c_{\gamma k\uparrow}\rangle\langle c^\dagger_{\beta j\downarrow}c_{\delta l\downarrow}\rangle\right)
\label{EDPairings}
\end{equation}
\end{widetext}
where $N_c=6$, $\alpha,\beta,\gamma,\delta =A,B$, $i,j,k,l= 1,2,3$ and $\lambda=1,...,9$. %The sums in $i,j$ run over the six unit cell sites while the sums in $\delta,\gamma$ run over the three possible nearest-neighbour sites of $i$ and $j$ respectively assuming periodic boundary conditions. 
The nine-component vector
$\Delta_\lambda$ contains the pairing solutions obtained from the linearized equations of Sec. \ref{sec:SCpairing} which are described in Appendix \ref{app::SCTc}. The correspondence between the vector components of ${\bf \Delta}$ and the 
Cooper paired bonds $(\alpha i, \beta j)$ are indicated in Eq. \eqref{eq:pairs}. As well as the expected value of the product of the creation and annihilation pairing operators, note that there is a substracted term which suppresses the contribution from the $i=j$ pairing amplitudes. Although these are substantial in small clusters, they are less important for quantifying actual superconducting correlations in the extended lattice.

Fig. \ref{fig::pairingcorrelations} illustrates the dependence of the triplet pairing correlations of the half-filled Hubbard model \eqref{eq:Hubbard} on a six-site cluster with $ J'/J \propto \lambda'^2/\lambda^2 $. The largest pairing correlations in a given channel, $\lambda$, 
$P_\lambda$, indicates that it is the most favorable at a given $J'/J$. The pairing correlations associated with the spatial symmetries A$_2$, B$_2$ and E$_2$ are degenerate in the whole $J'/J$ range explored due to the absence of intertriangle pairing amplitudes in these solutions making them indistinguishable on a six-site cluster. The
results of Fig. \ref{fig::pairingcorrelations} show that while the A$_2$, B$_2$ and E$_2$ solutions
are favored for $J'/J \leq 1/\sqrt{2}$, the $B_1$ followed by an $E_1^{}$ dominates for $J'/J > 1/\sqrt{2}$. This change of solutions with $J'/J$ is in overall qualitative agreement with the predictions of the linearized gap equations, Fig. \ref{fig:Tc}.

\section{Chern number in superconducting states}
\label{app::Chern}
In a discretized Brillouin zone, the total Chern number $\nu$ can be calculated from the Berry phases $\gamma_{nl}$ at each elementary placquette $l$. The Berry phase is  the accumulated phase of the wave function along a  closed $k$-path. 
\begin{equation}
\gamma_{nl}=\Im \ln\prod_{j=0}^{N-1}\bra{u_{\bf{nk}_j}}\ket{u_{\bf{nk}_{j+1}}}
\end{equation}
where $n$ is the band index and the loop chosen is rectangular with $N=4$. In the multiband case the wave functions overlap of all possible combinations must be taken into account. Thus we construct a $N_{c}\times N_{c}$ matrix at each step of the path, where  $N_{c}$  is
the number of occupied bands,. Then, the Berry phase is just the phase of the determinant of the product of these matrices along the loop. The Chern number is  the sum over the first Brillouin zone (FBZ) of all those Berry phases:
\begin{equation}
\nu=\frac{1}{2\pi }\sum_{FBZ} \Im\ln\det\prod_{j=0}^{3}\bra{u_{\bf{mk}_j}}\ket{u_{\bf{nk}_{j+1}}}
\label{ChernForm}
\end{equation}
where $1\leq m,n \leq N_c$. 

The Chern number $\nu$ of a 2-dimensional superconductor with broken time-reversal symmetry can be computed following the exact procedure stated above but replacing $\ket{u_{nk}}$ by the Bogoliubov quasiparticle wave functions.\cite{Trivedi}

\bibliography{biblio-3}

\end{document}